\documentclass[10pt,twocolumn,twoside]{IEEEtran}
\usepackage{cite}
\usepackage{amssymb}
\usepackage{float}
\usepackage{eqparbox}
\usepackage{multirow}
\usepackage{fixltx2e}
\usepackage{stfloats}
\usepackage[cmex10]{amsmath}
\usepackage{amsthm}

\usepackage{algorithm}
\usepackage{algorithmic}

\usepackage{array}

\usepackage{perpage}     
\MakePerPage{footnote} 

\usepackage{mdwmath}
\usepackage{mdwtab}
\usepackage{amsmath} \usepackage{amsfonts}
\makeatletter \def\amsbb{\use@mathgroup \M@U \symAMSb} \makeatother
\usepackage{bbold}
\usepackage{relsize} 
\usepackage{graphicx}
\usepackage{calc}
\usepackage{subfigure}
\usepackage[belowskip=-2pt,aboveskip=0pt]{caption}

\usepackage{amssymb}
\usepackage{eqparbox}
\usepackage{multirow}
\usepackage[cmex10]{amsmath}
\usepackage{fixltx2e}
\usepackage{stfloats}

\newcommand{\scrsmall}{\scriptscriptstyle }
\def\bphi{\mbox{\boldmath $\phi$}}
\def\bpsi{\mbox{\boldmath $\psi$}}

\def\bsigma{\mbox{\boldmath $\sigma$}}
\def\diag{\mbox{\rm{diag}}}
\def\col{\mbox{\rm{col}}}
\def\msd{{\rm{msd}}}
\def\emse{{\rm{emse}}}

\def\bvec{{\rm{bvec}}}

\def\Tr{{\rm{Tr}}}

\def\E{{\amsbb{E}}}
\def\u{\boldsymbol{u}}

\def\n{\boldsymbol{n}}

\def\v{\boldsymbol{v}}
\def\e{\boldsymbol{e}}
\def\f{\boldsymbol{f}}
\def\d{\boldsymbol{d}}
\def\z{\boldsymbol{z}}

\def\w{\boldsymbol{w}}

\def\be{\begin{equation}}
\def\ee{\end{equation}}
\def\ba{\begin{align}}
\def\ea{\end{align}}
\begin{document}
\title{Diffusion LMS Strategies in Sensor Networks with Noisy Input Data}
\author{Reza~Abdolee,~\IEEEmembership{Student Member,~IEEE,}
     and~ Benoit~Champagne,~\IEEEmembership{Senior Member,~IEEE,}%
\thanks{A short preliminary version of this work was presented in the European Signal Processing Conference (EUSIPCO), Aug. 2012 \cite{abdoleeDiffNoisyRegreesor2012}.} 
\thanks{R. Abdolee and B. Champagne are with the Department
of Electrical and Computer Engineering, McGill University, Montreal,
QC, H3A 0E9 Canada (e-mail: reza.abdolee@mail.mcgill.ca, benoit.champagne@mcgill.ca).}%
\thanks{This work was supported by the Natural Sciences and Engineering Research Council (NSERC) of Canada.}
\pagenumbering{gobble} 
}
\maketitle

\begin{abstract}
We investigate the performance of distributed least-mean square (LMS) algorithms for parameter estimation over sensor networks where the regression data of each node are corrupted by white measurement noise.  Under this condition, we show that the estimates produced by distributed LMS algorithms will be biased if the regression  noise is excluded from consideration. We propose a bias-elimination technique and develop a novel class of diffusion LMS algorithms that can mitigate the effect of regression noise and obtain an unbiased estimate of the unknown parameter vector over the network. In our development, we first assume that the variances of the regression noises are known \textit{a-priori}. Later, we relax this assumption by estimating these variances in real-time.
We analyze the stability and convergence of the proposed algorithms and derive closed-form expressions to characterize their mean-square error performance in transient and steady-state regimes.
We further provide computer experiment results that illustrate the efficiency of the proposed algorithms and support the analytical findings.
\end{abstract}
\IEEEpeerreviewmaketitle
\section{Introduction}
\label{sec:intro}
\IEEEPARstart{O}NE of the critical issues encountered in distributed parameter estimation over sensor networks is the distortion of the collected regression data by noise, which
occurs when the local copy of the underlying system input signal at each node is corrupted by various sources of impairments such as measurement or quantization noise.
This problem has been extensively investigated for the case of single-node processing devices 
\cite{golub1980analysis,van1991total,degroat1993data,regalia1994unbiased,ho1995bias, shin2001bias,so1999modified,davila1994efficient, feng1998total,so2001lms,jo2005consistent,sagara1977line,jia2001bias,feng2000modified,zheng1989unbiased,kang2013bias}. These studies have shown that if the deleterious effect of the input noise is not taken into account, the parameter estimates so obtained will be inaccurate and biased.
Various practical solutions have been suggested to mitigate the effect of the input measurement noise or to remove the bias from the resulting estimates 
\cite{regalia1994unbiased,ho1995bias,shin2001bias,so1999modified,davila1994efficient, feng1998total,so2001lms,jo2005consistent,sagara1977line,jia2001bias,feng2000modified,zheng1989unbiased,kang2013bias}.
These solutions, however, may no longer leads to optimal results in sensor networks with decentralized processing structure where the data measurement and parameter estimation are performed at multiple processing nodes in parallel and with cooperation. 

For networking applications, a distributed total-least-squares (DTLS) algorithm has been proposed that is developed using semidefinite relaxation and convex semidefinite programming \cite{bertrand2011consensus}. 
This algorithm mitigates the effect of white input noise by running a local TLS algorithm at each sensor node and exchanging the locally estimated parameters between the nodes for further refinement.
The DTLS algorithm computes the eigendecomposition of an augmented covariance matrix at every iteration for all nodes in the network, and is therefore mainly suitable for applications involving nodes with powerful processing abilities. 
In a follow up paper, the same authors proposed a low-complexity DTLS algorithm \cite{bertrand2012low} that uses an inverse power iteration technique to reduce the computational complexity of the DTLS while demanding lower communication power. 

In recent years, several classes of distributed \emph{adaptive} algorithms for parameter estimation over networks have been proposed, including incremental \cite{bertsekas1997new,Nedia2001Incremental,rabbat2005quantized,lopes2007incremental}, consensus \cite{tsitsiklis1986distributed,xiao2006space,braca2008running,sardellitti2010fast,dimakis2010gossip,nedic2009distributed,sardellitti2010fast,kar2011convergence}, and diffusion algorithms \cite{lopes2008diffusion,cattivelli2008diffusion,cattivelli2010diffusion,di2011bio,takahashi2010diffusion, chouvardas2011adaptive,chen2012diffusion,tu2012diffusion,zhao2012performance,Sayed2013diffusion, di2013sparse, tu2014decision}. 
Incremental techniques require the definition of a cyclic path over the nodes, which is generally an NP-hard problem; these techniques are also sensitive to link failures.
Consensus techniques require doubly-stochastic combination policies and, when used in the context of adaptation with constant step-sizes, can lead to unstable behavior even if all individual nodes can solve the inference task in a stable manner \cite{tu2012diffusion}.
In this work, we focus on diffusion strategies because they have been shown to be more robust and to lead to a stable behavior regardless of the underlying topology, even when some of the underlying nodes are unstable \cite{tu2012diffusion}.

A bias-compensated diffusion-based recursive least-squares (RLS) algorithm has been developed in \cite{bertranddiffusion2011} that can obtain unbiased estimates of the unknown system parameters over sensor networks, where the regression data are distorted by colored noise.  While this algorithm offers fast convergence speed, its high computational complexity and numerical instability may be a hindrance in some applications.
In contrast, the diffusion LMS algorithms are characterized by low complexity and 
numerical stability. Motivated by these features, in this paper, we investigate the performance of standard diffusion LMS algorithms \cite{lopes2008diffusion,cattivelli2008diffusion,cattivelli2010diffusion} over sensor networks where the input regression data are corrupted by additive white noise. To overcome the limitations of these algorithms, as exposed by our analysis under this scenario, we then propose an alternative problem formulation that leads to a novel class of diffusion LMS algorithms, which we call bias-compensated diffusion strategies.

More specifically, we first show that in the presence of noisy input data, the parameter estimates produced by standard diffusion LMS algorithms are biased. We then reformulate this estimation problem in terms of an alternative cost function and develop bias-compensated diffusion LMS strategies that can produce unbiased estimates of the system parameters. The development of these algorithms relies on a bias-elimination strategy that assumes prior knowledge about the regression noise variances over the network. The analysis results show that if the step-sizes are within a given range, the algorithms will be stable in the mean and mean-square sense and the estimated parameters will converge to their true values. Finally, we relax the known variance assumption by incorporating a recursive approach into the algorithm to estimate the variances in real-time.

In summary, the contributions of this article are: 
a) performance evaluation of standard diffusion LMS algorithms in networks with noisy input regression data; 
b) development of a novel class of diffusion LMS strategies that are robust under this condition;
c) presentation of a recursive estimation approach to obtain the regression noise variances without using the second order statistics of the data;
d) derivation of conditions under which the proposed algorithms are stable in the mean and mean-square sense;
e) characterization of their mean-square deviation (MSD) and excess mean-square error (EMSE) in transient and steady-state regimes; and
f) validation of theoretical findings through numerical simulations of newly proposed algorithms for parameter estimation over sensor networks.

The remainder of the paper is organized as follows. In the next section, we formulate the problem and discuss the effects of input measurement noise on the performance of diffusion LMS over sensor networks. In Section \ref{sec.:UnbiasedCentralizedLms}, we propose bias-compensated diffusion LMS algorithms along with a recursive estimation of the regression noise variance. In Section \ref{sec.:UnbiasedDiffLmsAnalysis}, we analyze the stability and convergence behavior of the developed algorithms, and obtain conditions under which the algorithms are stable in the mean and mean-square sense.
We present the computer experiment results in Section \ref{sec.:UnbiasedDiffLmsResults}, and conclude the paper in Section \ref{sec.:UnbiasedDiffLMSConclusion}.

\textit{Notation:} Matrices are represented by uppercase fonts, vectors  by lowercase fonts. Boldface letters are reserved for random variables and normal letters are used for deterministic variables. Superscripts $(\cdot)^T$  and $(\cdot)^{\ast}$, respectively, denote transposition and conjugate transposition. Symbols $\Tr(\cdot)$ and $\rho(\cdot)$ denote the trace and spectral radius of their matrix argument. The operator $\E[\cdot]$ stands for statistical expectation, and $\lambda_k(\cdot)$ denotes the $k$-th eigenvalue of its matrix argument. The Kronecker product  is denoted by $\otimes$, and the \textit{block Kronecker products} \cite{koning1991block} is denoted by $\otimes_{\scrsmall{b}}$. The operator diag$\{\cdot\}$ converts its argument list into a (block) diagonal matrix. The operator col$\{\cdot\}$ performs a vertical stacking of its arguments while vec$(\cdot)$ is the standard vectorization for matrices. The symbol bvec($\cdot$) is the block vectorization operator that transforms a block-partitioned matrix into a column vector  \cite{koning1991block}.

\section{Problem Statement}
\label{sec.:problemStatement}
\begin{figure}[h!]
\centering
\includegraphics[scale=0.75]{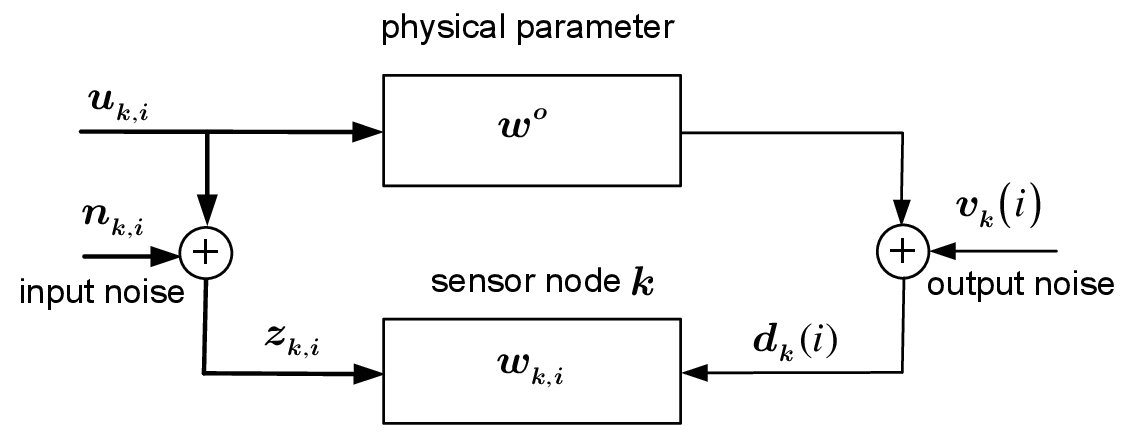}
\caption{Measurement model for node $k$.}
\label{fig.:sensory_regressor_system_model}
\end{figure}

Consider a collection of $N$ sensor nodes distributed over a geographical area and used to monitor a physical phenomenon characterized by some unknown parameter vector
$w^o \in {\amsbb C}^{M \times 1}$.
As illustrated in Fig. \ref{fig.:sensory_regressor_system_model}, at discrete-time $i \in {\amsbb N}$, each node $k \in \{1,2,\cdots,N\}$ collects noisy samples of the system input and output denoted by $\z_{k,i}\in {\amsbb C}^{1\times M}$ and $\d_k(i)\in {\amsbb C}$, respectively. These measurement samples can be expressed as:
\begin{align}
\z_{k,i}&= \u_{k,i}+ \n_{k,i}
\label{eq.:network_linear_model_eq1}\\
\d_k(i)&= \u_{k,i} w^o+ \v_{k}(i) \label{eq.:network_linear_model_eq2}
\end{align}
where $\u_{k,i} \in {\amsbb C}^{1 \times M}$,  $\n_{k,i}\in {\amsbb C}^{1 \times M}$, and $\v_{k}(i) \in {\amsbb C}$, respectively, denote the regression data vector, the input measurement noise vector, and the output measurement noise\footnote{We use parentheses to refer to the time indices of scalar variables, such as $\d_k(i)$, and subscripts to refer to time indices of vector variables, such as $\z_{k,i}$.}.
\newtheorem{assump}{Assumption}
\begin{assump}
\label{assump.:data model}
The random variables in data model (\ref{eq.:network_linear_model_eq1})-(\ref{eq.:network_linear_model_eq2}) satisfy the following conditions:
\begin{itemize}
  \item[a)] The regression data vectors are independent and identically distributed (i.i.d.) over time and independent over space, with zero-mean and covariance matrix \mbox{$R_{u,k}=\E[\u_{k,i}^* \u_{k,i}]>0$}.
  \item[b)] The regression noise vectors $\n_{k,i}$ are Gaussian, i.i.d. over time and independent over space, with zero-mean and covariance matrix \mbox{$R_{n,k}=\E[\n_{k,i}^* \n_{k,i}]=\sigma^2_{n,k} I$}.
  \item[c)]  The output noise samples $\v_k(i)$ are i.i.d. over time and independent over space, with zero-mean and variance $\sigma^2_{v,k}$.
  \item[d)] The random variables $\u_{k,i}$, $\n_{\ell,j}$ and $\v_p(m)$ are independent for all $k,\ell,p, i, j$, and $m$.
\end{itemize}
\end{assump}
\vspace{0.5cm}
\par \noindent
The linear model (\ref{eq.:network_linear_model_eq1})-(\ref{eq.:network_linear_model_eq2}) differs from those used in previous works on distributed estimation, such as \cite{lopes2007incremental,lopes2008diffusion, cattivelli2010diffusion}.
In these references, it is assumed that the actual regression vector $\u_{k,i}$ is available at each node $k$. There are many practical situations, however, where the nodes only have access to noisy measurements of the regression data.
We use relation (\ref{eq.:network_linear_model_eq1}) to model such disturbance in the regressors, and to investigate the effect of the noise process $\n_{k,i}$ on the distributed estimation of $w^o$.
To better understand the effect of this noise, we first examine the behavior of a centralized estimation solution under this condition and then explain how the resulting effect carries over to distributed approaches.

In centralized estimation, nodes transmit their measurement data $\{\z_{k,i}, \d_k(i)\}_{k=1}^N$ to a central processing unit.
In the absence of measurement noise,
\mbox{i.e., $\n_{k,i}=0$}, the central processor can estimate the unknown parameter vector $w^o$  by, e.g., minimizing the following mean-square error (MSE) function \cite{sayed2008}:
\be
J_u(w)=\sum _{k=1}^N \E|\d_{k}(i)-\u_{k,i} w|^2.
\label{eq.noiselessObjFunction}
\ee
Let us introduce $r_{du,k}\triangleq \E[\d_k(i) \u_{k,i}^*]$ and denote the sums of covariance matrices and cross-covariance vectors over the set of nodes by:
\begin{equation}
 R_{u}=\sum_{k=1}^N R_{u,k},\quad r_{du}=\sum_{k=1}^N r_{du,k}.
\end{equation}
It can be verified that under Assumption \ref{assump.:data model}, the solution of (\ref{eq.noiselessObjFunction}) is:
\begin{equation}
 w^o=R_{u}^{-1}\,r_{du}.
\label{eq.:SolutionUnbiasCentralized}
\end{equation}
Now consider the recovery of the unknown parameter vector $w^o$ for the noisy regression system described by (\ref{eq.:network_linear_model_eq1}) and (\ref{eq.:network_linear_model_eq2}).
Since the regression noise $\n_{k,i}$ is independent of $\u_{k,i}$ and $\d_k(i)$, we have
\begin{align}
R_{z,k}&\triangleq \E[\z_{k,i}^* \z_{k,i}]=R_{u,k}+\sigma^2_{n,k} I \\
 r_{dz,k}&\triangleq \E[\d_{k}(i) \z_{k,i}^*]=r_{du,k}.
\end{align}
Considering these relations and now minimizing the global MSE function
\be
J_z(w)=\sum _{k=1}^N \E|\d_{k}(i)-\z_{k,i} w|^2
\label{eq.noisyObjFunction}
\ee
with $\u_{k,i}$ in (\ref{eq.noiselessObjFunction}) replaced by $\z_{k,i}$ in (\ref{eq.noisyObjFunction}), we arrive at the biased solution
\begin{equation}
 w^b=\big(R_{u}+\sigma^2_{n} I \big)^{-1} r_{du}
\label{eq.:SolutionBiasCentralized}
\end{equation}
where
\be
\sigma^2_{n}=\sum_{k=1}^N \sigma^2_{n,k}.
\ee

Let us define the bias implicit in solution (\ref{eq.:SolutionBiasCentralized}) as \mbox{$b=w^o-w^b$}. To evaluate $b$, we may use the following identity, which holds for square matrices $X_1$ and $X_2$ provided that $X_1$ and $X_1+X_2$ are both invertible \cite{miller1981inverse}:
\be
(X_1+X_2)^{-1}=X_1^{-1}-(I+X_1^{-1} X_2)^{-1}X_1^{-1} X_2 X_1^{-1}.
\ee
Here $R_{u}$ and $(R_{u}+\sigma^2_{n}\, I)$ are invertible, and therefore, we obtain:
\be
\big(R_{u}+\sigma^2_{n}\,I\big)^{-1}=R_{u}^{-1}-\sigma^2_{n}(I+\sigma^2_{n}R_{u}^{-1})^{-1}R_{u}^{-2}.
\ee
Considering this expression and relation (\ref{eq.:SolutionBiasCentralized}),
the bias resulting from the minimum MSE estimation at the fusion center can be expressed as:
\begin{equation}
b=\sigma^2_{n}(I+\sigma^2_{n} R_{u}^{-1})^{-1}R_{u}^{-1}w^o.
\label{eq.:bias-term}
\end{equation}

In the absence of regressor noise, it has been shown in previous studies that the parameter estimates obtained from standard diffusion LMS strategies approach the minimizer of the network global MSE function \cite{cattivelli2010diffusion}.
This also holds in noisy regression applications for diffusion LMS developed based on the global cost (\ref{eq.noisyObjFunction}), meaning that the estimates generated by standard diffusion LMS algorithms will eventually approach (\ref{eq.:SolutionBiasCentralized}). As shown by (\ref{eq.:bias-term}), this solution is biased and deviates from the optimal estimate by $b$.
This issue will become more explicit in our convergence analysis in Section \ref{subsec:mean-analysis}.

In sequel, we explain how by forming a suitable objective function, the bias can be compensated in both centralized and distributed LMS implementations.

\section{Bias-Compensated LMS Algorithms}
\label{sec.:UnbiasedCentralizedLms}
In our development, we initially assume that the regression noise variances, $\{\sigma^2_{n,k}\}_{k=1}^N$, are known \textit{a-priori}. We later remove this assumption by estimating these variances in real-time.  In networks with centralized signal processing structure, one way to obtain the unbiased optimal solution (\ref{eq.:SolutionUnbiasCentralized}) is to search for
a global cost function whose gradient vector is identical to that of cost (\ref{eq.noiselessObjFunction}).
It is straightforward to verify that the following global cost function satisfies this requirement:
\be
J(w)=\Big(\sum _{k=1}^N \E|\d_{k}(i)-\z_{k,i} w|^2\Big)-\Big(\sum _{k=1}^N \sigma^2_{n,k} \|w\|^2\Big).
\label{eq.:modified-global-objective function}
\ee
\newtheorem{Remark}{Remark}
\begin{Remark}
In bias-compensation techniques for single-node adaptive algorithms, including \cite{zheng1989unbiased,sagara1977line,jo2005consistent}, the authors first apply a least squares (LS) or minimum MSE procedure to obtain an estimate of the unknown parameter vector. The resulting estimate consists of the desired solution along with an additive bias term.
The bias, which is normally expressed in terms of the second order statistics of the
regression data and the input and output measurement noises, is removed from the solution by subtraction.
In the proposed technique in this paper, we start by considering bias removal one step earlier, meaning that we design a convex objective function such that its unique stationary point leads to an unbiased estimate.
From this respect, our approach is mostly inspired from the derivation of the modified LMS and RLS algorithms in \cite{so1999modified,feng2000modified}. However, these algorithms still assume the knowledge of the ratio of input-to-output noise variances in their update equations.
\end{Remark}

The derivation of distributed algorithms will be made easier if we can decouple the network global cost function and write it as sum of local cost functions that are formed using the local data. The global cost (\ref{eq.:modified-global-objective function}) already has such a desired form. For this to become more explicit, we express  (\ref{eq.:modified-global-objective function}) as:
\begin{align}
J(w)=\sum_{k=1}^N J_k( w)
\label{eq.:modified-global-objective function2}
\end{align}
where $J_k( w)$ is the cost function associated with node $k$ and is
given in terms of local data $\d_k(i)$ $\z_{k,i}$, i.e.,
\begin{align}
J_k( w)=\E| \d_k(i)- \z_{k,i} w|^2-\sigma^2_{n,k}\| w\|^2.
\label{eq.:modified-cost-node-k}
\end{align}
\par \noindent
\begin{Remark}
\label{re:Jk-positive-definite}
Under Assumption \ref{assump.:data model}, the Hessian matrix of (\ref{eq.:modified-cost-node-k}) is positive definite, i.e., $\nabla^2_w J_k(w)>0$, hence,  $J(w)$ is strongly convex \cite{boyd2004convex}.
\end{Remark}

Below, we first comment on the centralized LMS algorithm that solves (\ref{eq.:modified-global-objective function}), and then elaborate on how to develop the unbiased distributed counterparts.
\subsection{Bias-Compensated Centralized LMS Algorithm}
To minimize (\ref{eq.:modified-global-objective function2}) iteratively, a centralized steepest descent algorithm \cite{sayed2008} can be implemented as:
\be
\w_i= \w_{i-1}-\mu \Big[\sum_{k=1}^N \nabla  J_k(\w_{i-1}) \Big]^*
\ee
where $\mu>0$ is the step-size, and $\nabla J_k(w)$ is a row vector representing the gradient of $J_k$ with respect to the vector $w$.
Computing the gradient vectors from (\ref{eq.:modified-cost-node-k}) leads to:
\begin{align}
\w_i= \w_{i-1}+\mu\sum_{k=1}^N \Big(r_{dz,k}-R_{z,k} \w_{i-1}+\sigma^2_{n,k}\w_{i-1}\Big).
\label{eq.:steepest-decsent-biasedCompensated}
\end{align}
In practice, the moments $R_{z,k}$ and $r_{dz,k}$ are usually unavailable. We, therefore, replace these moments by their instantaneous approximations $\z_{k,i}^* \z_{k,i}$ and $\z_{k,i}^*  \d_k(i)$, respectively, and obtain the \mbox{bias-compensated centralized LMS} algorithm:
\be
\w_i=\w_{i-1}+\mu\sum_{k=1}^N \Big(\z_{k,i}^*[\d_k(i)-\z_{k,i}\w_{i-1}]+\sigma_{n,k}^2\w_{i-1}\Big).
\ee
In Section \ref{subsec.:RegressorNoiseVarianceEstimation}, we propose an adaptive scheme to estimate the variances of the regression noise required in the above centralized LMS algorithm as well as in its distributed counterpart derived below.
\subsection{Bias-Compensated Diffusion LMS Strategies}
\label{sec.:UnbiasedDiffLms}
There exist different distributed optimization techniques that can be applied on (\ref{eq.:modified-global-objective function}) to find  $w^o$  \cite{bertsekas1989parallel,lopes2008diffusion,cattivelli2010diffusion}. We concentrate on diffusion strategies \cite{lopes2008diffusion,cattivelli2010diffusion} because they endow the network with real-time adaptation and learning abilities.
In particular, diffusion optimization strategies lead to distributed algorithms that can estimate the parameter vector $w^o$ and track its changes over time \cite{lopes2008diffusion,cattivelli2010diffusion,chen2012diffusion,sayed2012diffusion}. Here, we briefly explain how diffusion LMS algorithms can be developed for parameter estimation in systems with noisy regression data. The main step in the development of these algorithms is to reformulate the global cost (\ref{eq.:modified-global-objective function}) and represent it as a group of optimization problems of the form:
\begin{align}
\min_w\Bigg\{ &\sum_{\ell\in {\cal N}_k} c_{\ell,k}\Big(\E| \d_{\ell}(i)- \z_{\ell,i} w|^2 -\sigma^2_{n,\ell}
\| w\|^2 \Big)\nonumber \\
&\hspace{3cm}+\sum_{\ell\in {\cal N}_k\backslash \{k\}} b_{\ell,k} \|w-w^o\|^2\Bigg\}.
\label{eq.modified-local-objective-function}
\end{align}
where ${\cal N}_k$ is the set of nodes with which node $k$ shares information, including node $k$ itself.
The nonnegative scalars $\{c_{\ell,k}\}$  are the entries of a right-stochastic matrix $C\in {\amsbb R}^{N\times N}$ which satisfy
\vspace{-0.2cm}
\be
c_{\ell,k}=0 \; {\rm if}\; {\ell} \notin \mathcal{N}_k,\; {\rm and}\;\;  \sum_{k=1}^N c_{\ell,k}=1.
\label{eq.:c-properties}
\ee
The scalars $\{b_{\ell,k}\}$ are scaling coefficients that will end up being incorporated into the combination coefficients $\{a_{\ell,k}\}$ that appear in the final statement (\ref{eq.:ATC-SecondStatement}) of the algorithm below. The first term in the objective function (\ref{eq.modified-local-objective-function}) is the modified mean-squared function incorporating the noise variances of neighboring nodes $\ell \in {\cal N}_k$. 
This part of the objective is based on the same strategy as in the above centralized objective function for bias removal. The second term in (\ref{eq.modified-local-objective-function}) is in fact a constraint that forces the estimate of the node $k$ to be aligned with the true parameter vector $w^o$. Since $w^o$ is not known initially, it will be alternatively substituted by an appropriate vector during the optimization process.
One can use the cost function (\ref{eq.modified-local-objective-function}) and follow similar arguments to those used in \cite{cattivelli2010diffusion,chen2012diffusion,sayed2012diffusion} to arrive at the bias-compensated adapt-then-combine (ATC) LMS strategy (Algorithm \ref{alg.:ATC Bias-Compensated Diffusion LMS}). Due to space limitations, these steps are omitted.
%
\begin{algorithm}
\caption{:\, ATC Bias-Compensated Diffusion LMS}
\label{alg.:ATC Bias-Compensated Diffusion LMS}
\begin{align}
{\bpsi}_{k,i}&=\w_{k,i-1}-\mu_k\displaystyle\sum_{\ell\in \mathcal{N}_k}c_{\ell,k} \big [\widehat{\nabla  J_{\ell}}(\w_{k,i-1})\big]^*  \\
\w_{k,i}&=\displaystyle\sum_{\ell\in \mathcal{N}_k}a_{\ell,k} {\bpsi}_{\ell,i} \label{eq.:ATC-SecondStatement}
\end{align}
\end{algorithm}
\par \noindent
In this algorithm, $\mu_k>0$ is the step-size at node $k$, the vectors $\bpsi_k$ and $\w_{k,i}$ are the intermediate estimates of $w^o$ at node $k$, and the stochastic gradient vector is computed as:
\begin{align}
\big[\widehat{\nabla  J_{\ell}}(\w_{k,i-1})\big]^*=&- \,\big[ \z_{\ell,i}^{\ast} \big(\d_{\ell}(i)- \z_{\ell,i} \w_{k,i-1}\big) \nonumber \\
&\hspace{2.5cm}+{\sigma}^2_{n,\ell}\w_{k,i-1}\big].
\end{align}
which is an instantaneous approximation to gradient of (\ref{eq.:modified-cost-node-k}).
Moreover, the nonnegative coefficients $a_{\ell,k}$ are the elements of a left-stochastic matrix $A \in {\amsbb R}^{N\times N}$ satisfying
\be
a_{\ell,k}=0\; {\rm if} \; {\ell} \notin \mathcal{N}_k, \; {\rm and}\; \sum_{\ell\in {\mathcal N}_k} a_{\ell,k}=1.
\label{eq.:a-properties}
\ee
To run the algorithm, we only need to select the coefficients $\{c_{\ell,k},a_{\ell,k}\}$, which can be computed based on any combination rules that satisfy (\ref{eq.:c-properties}) and (\ref{eq.:a-properties}).
One choice to compute the entries of matrix $A$ is:
\begin{equation}
a_{\ell,k}=\frac{\sigma^{-2}_{n,\ell}}{\sum_{\ell\in {\mathcal N}_k}\sigma^{-2}_{n,\ell}}\quad {\rm and} \quad a_{k,k}=1-\sum_{\ell\in {\mathcal N}_k \backslash k} a_{\ell,k}.
\label{eq.:relative_variance}
\end{equation}
This rule implies that the entry $a_{\ell,k}$ is inversely proportional to the regressor noise variance of node $\ell$. Other left-stochastic choices for $A$ are possible, including those that take into account both the noise variances and the degree of connectivity of the nodes \cite{zhao2012performance}.

By reversing the order of the adaptation and combination steps in Algorithm \ref{alg.:ATC Bias-Compensated Diffusion LMS}, we can obtain the following combine-then-adapt (CTA) diffusion strategy.
%
\begin{algorithm}
\caption{:\, CTA Bias-Compensated Diffusion LMS}
\label{alg.:CTA Bias-Compensated Diffusion LMS}
\begin{align}
&\bpsi_{k,i-1}=\displaystyle\sum_{\ell\in {\cal N}_k}a_{\ell,k} \w_{\ell,i-1}\\
&{\w}_{k,i}=\bpsi_{k,i-1}-\mu_k\displaystyle\sum_{\ell\in \mathcal{N}_k} c_{\ell,k} \big[\widehat{\nabla  J_{\ell}}(\bpsi_{k,i-1})\big]^*
\end{align}
\end{algorithm}
As we will show in the analysis, the proposed ATC and CTA bias-compensated diffusion-LMS, in average, will converge to the unbiased solution (\ref{eq.:SolutionUnbiasCentralized}) even when the regression data are corrupted by noise. In comparison, the estimate of the previous diffusion LMS strategies such as one proposed in \cite{cattivelli2010diffusion} will be biased under such condition.

\begin{Remark}
In the proposed ATC algorithm, each node $k$ receives $\{\u_{\ell,i}, \d_{\ell}(i), \sigma^2_{n,\ell}\}$ from its neighbors in the adaptation step, and $\bpsi_{\ell,i}$ in the combination step, where $\ell \in {\cal N}_k$. In total, it will receive $(2M+2)|{\cal N}_k|$ scalar data from its neighbors. To reduce the communication overhead of the network, one solution is to choose $C=I$. Doing so, we can reduce the amount of exchanged data at each node $k$ to $M |{\cal N}_k|$ while maintaining almost similar performance results, as evidenced in Section \ref{sec.:UnbiasedDiffLmsResults}. Note that the amount of information exchange in this case will be equal to that of the standard ATC diffusion LMS in \cite{cattivelli2010diffusion}. This conclusion is also valid for the proposed CTA Algorithm \ref{alg.:CTA Bias-Compensated Diffusion LMS}.
\end{Remark}

\section{Performance Analysis}
\label{sec.:UnbiasedDiffLmsAnalysis}
In this section, we analyze the convergence and stability of the proposed ATC and CTA bias-compensated diffusion LMS algorithms by viewing them as special cases of a more general diffusion algorithm of the form:
\begin{align}
&{\bphi}_{k,i-1}=\displaystyle\sum_{\ell\in {\cal N}_k}a^{(1)}_{\ell,k} \w_{\ell,i-1} \label{eq.update 1}\\
&{\bpsi}_{k,i}=\bphi_{k,i-1}-\mu_k\sum_{\ell\in {\cal N}_k} c_{\ell,k}\big[\widehat{\nabla_{\phi} J_{\ell}}(\bphi_{k,i-1)}\big]^*\label{eq.update 2}\\
&\w_{k,i}=\displaystyle\sum_{\ell\in {\cal N}_k}a^{(2)}_{\ell,k} \bpsi_{\ell,i} \label{eq.update 3}
\end{align}
where $\{a^{(1)}_{\ell,k}\}$ and  $\{a^{(2)}_{\ell,k}\}$ are non-negative real coefficients corresponding to the $(\ell,k)$-th entries of left-stochastic matrices $A_1$ and $A_2$, respectively, which have the same properties as $A$. Different choices for $A_1$ and $A_2$ corresponds to different operation modes. For instance, $A_1=I$ and $A_2=A$ correspond to ATC whereas $A_1=A$ and $A_2=I$ generate CTA. For mathematical tractability, in our analysis, we assume that the variances of the regression noises, i.e., $\sigma_{n,k}^2$, over the network are known \textit{a-priori}.

We define the local weight-error vectors as $\tilde \w_{k,i}= w^o- \w_{k,i}$, $\tilde{\bpsi}_{k,i}= w^o-\bpsi_{k,i}$ and $
\tilde{\bphi}_{k,i}= w^o-\bphi_{k,i}$, and form the global weight-error vectors, by stacking the local error vectors, i.e.:
\begin{align}
&\tilde{\bphi}_i=\col\{\tilde{\bphi}_{1,i},\tilde{\bphi}_{2,i},\ldots,\tilde{\bphi}_{N,i}\}\\
&\tilde{\bpsi}_i=\col\{\tilde{\bpsi}_{1,i},\tilde{\bpsi}_{2,i},\ldots,\tilde{\bpsi}_{N,i}\}\\
&\tilde \w_i=\col\{\tilde \w_{1,i},\tilde \w_{2,i},\ldots,\tilde \w_{N,i}\}.
\end{align}
We also define the block variables:
\small
\begin{align}
&\boldsymbol{g}_i={\mathcal C}^T \text{col}\{ \z_{\ell,i}^{\ast} \v_1(i),\ldots, \z_{N,i}^{\ast} \v_N(i) \} \label{eq.:Gi-deginition}\\
&\boldsymbol{\mathcal{R}}_i=\diag\Big \{ \sum_{\ell \in {\cal N}_k}c_{\ell,k} \,( \z_{\ell,i}^{\ast} \z_{\ell,i}-\sigma^2_{n,\ell}I), k=1,\cdots,N \Big\}
\label{eq.:Ri}\\
&\boldsymbol{\mathcal {P}}_i=\diag\Big \{ \sum_{\ell \in {\cal N}_k}c_{\ell,k} \,( \z_{\ell,i}^{\ast} \n_{\ell,i}-\sigma^2_{n,\ell}I), k=1,\cdots,N \Big\} \label{eq.:Pi}\\
&{\mathcal M}=\diag \big \{ \mu_1 I_M,\cdots, \mu_N I_M \big \}
\end{align}
\normalsize
and introduce the following extended combination matrices:
\begin{align}
{\mathcal A_1}=A_1\otimes I_M, \quad {\cal A}_2=A_2\otimes I_M, \quad {\mathcal C}=C\otimes I_M.
\end{align}
Using these definitions and update equations (\ref{eq.update 1})-(\ref{eq.update 3}), it can be verified that the following relations hold:
\begin{align}
&\tilde{\bphi}_{i-1}={\cal A}_1^T\tilde \w_{i-1} \nonumber \\
&\tilde{\bpsi}_i=\tilde{\bphi}_{i-1}-{\mathcal M}( \boldsymbol{g}_i-\boldsymbol{\mathcal {P}}_i \omega^o+\boldsymbol{\mathcal{R}}_i \tilde{\bphi}_{i-1}) \nonumber \\
&\tilde \w_i={\cal A}_2^T\tilde{\bpsi}_i
\label{eq.global_error_vector_set}
\end{align}
where $\omega^o={\mathbb 1} \otimes w^o$. From the set of equations given in (\ref{eq.global_error_vector_set}), it is deduced that the network error vector $\tilde{\w}_i$ evolves with time according to
the recursion:
\begin{align}
\tilde \w_i=\boldsymbol{\cal B}_i\tilde \w_{i-1}-{\cal A}_2^T{\mathcal M} \boldsymbol{{g}}_i+{\cal A}_2^T{\mathcal M}\boldsymbol{\mathcal {P}}_i \omega^o
\label{eq.:global_error_vector_w}
\end{align}
where the time-varying matrix $\boldsymbol{\cal B}_i$ is defined as:
\begin{equation}
\boldsymbol{\cal B}_i={\cal A}_2^T(I-{\mathcal M}\boldsymbol{\cal{R}}_i){\cal A}^T_1.
\end{equation}
\subsection{Mean Convergence and Stability}
\label{subsec:mean-analysis}
Tacking the expectation of both sides of
(\ref{eq.:global_error_vector_w}) and considering Assumption \ref{assump.:data model}, we arrive at:
\begin{align}\
\E[\tilde \w_i]= {\cal B} \Big (\E [\w_{i-1}]\Big)
\label{eq.:mean_perfomance}
\end{align}
where in this relation:
\small
\begin{align}
&{\cal B}\triangleq\E [\boldsymbol{{\cal B}}_i]={\cal A}_2^T(I-{\mathcal M}{\mathcal R}){\mathcal A_1}^T\\
&{\mathcal R}\triangleq\E [\boldsymbol{\mathcal{R}}_i]= \diag\Big \{ \sum_{\ell\in {\cal N}_k} c_{\ell,k} \, R_{u,\ell}, k=1,\cdots, N \Big \}.
\end{align}
\normalsize
To obtain (\ref{eq.:mean_perfomance}), we used the fact that $\E[{\mathcal
A_2}^T{\mathcal M} \boldsymbol{g}_i]=0$ because $ \v_{k,i}$ is independent of $\z_{k,i}$ and $\E[ \v_{k}(i)]=0$. Moreover, we have $\E[\boldsymbol{\mathcal {P}}_i]=0$ because $\E[\z_{\ell,i}^{\ast} \n_{\ell,i}]=\sigma^2_{n,\ell}I$. According to (\ref{eq.:mean_perfomance}), $\lim_{i \rightarrow \infty} \E\|\tilde \w_i\|\rightarrow 0$ if
${\cal B}$ is stable ( i.e., when $\rho({\cal B})<1)$. In fact, because $\rho({\mathcal A_1})=\rho({\cal A}_2)=1$ and ${\mathcal R}>0$ choosing the step-sizes according to:
\begin{equation}
0<\mu_k <\frac{2}{\rho\big( \sum_{\ell \in {\cal N}_k} c_{\ell,k}{R_{u,\ell}}\big)}
\label{eq.:mu_range}
\end{equation}
guarantees $\rho({\cal B})<1$. We omit the proof. The similar argument can be found in \cite{sayed2012diffusion} and \cite{takahashi2010diffusion}. We summarize the mean-convergence results of the proposed bias-compensated diffusion LMS in the following.
\newtheorem{thm}{Theorem}
\begin{thm}
\label{thm.:UnbiasedDLMSmeanConvergence}
Consider an adaptive network that operates using diffusion Algorithms \ref{alg.:ATC Bias-Compensated Diffusion LMS} or \ref{alg.:CTA Bias-Compensated Diffusion LMS} with the space-time data (\ref{eq.:network_linear_model_eq1}) and (\ref{eq.:network_linear_model_eq2}).
In this network, if we assume that the regressors noise variances are known or perfectly estimated, the mean error vector evolves with time according to (\ref{eq.:mean_perfomance}). Furthermore, Algorithms \ref{alg.:ATC Bias-Compensated Diffusion LMS} and \ref{alg.:CTA Bias-Compensated Diffusion LMS} will be asymptotically unbiased and stable provided that the step-sizes satisfy (\ref{eq.:mu_range}).
\end{thm}
%
\begin{Remark}
In networks with noisy regression data (\ref{eq.:network_linear_model_eq1}), the estimates generated by the previous diffusion LMS strategies such as the ones proposed in \cite{cattivelli2010diffusion,sayed2012diffusion} are biased, i.e, $\E[\tilde \w_i]\neq 0$ as $i \rightarrow \infty$. This can be readily shown if we remove \mbox{$\sigma^2_{n,k}$} from (\ref{eq.:Ri}) and (\ref{eq.:Pi}). In this scenario,  (\ref{eq.:mean_perfomance}) will be stable if
\begin{equation}
0<\mu_k <\frac{2}{\rho\Big( \sum_{\ell \in {\cal N}_k}{ c_{\ell,k} \big(R_{u,\ell}}+\sigma^2_{n,\ell}I_M\big) \Big)}.
\label{eq.:mu_range_biased}
\end{equation}
Then, for sufficiently small step-sizes, satisfying  (\ref{eq.:mu_range_biased}), it can be verified that the estimate of the standard diffusion LMS deviates from the network optimal solution $\omega^o$ by: 
\begin{align}
\lim_{i\rightarrow \infty}\E[\tilde \w_i]= (I_{NM}-{\cal B}')^{-1}{\cal A}_2^T{\mathcal M} {\cal P}' \omega^o
\label{eq.:mean_perfomance-biased}
\end{align}
where
\begin{align}
{\cal B}'&\triangleq {\cal A}_2^T(I_{NM}-{\mathcal M}{\cal R}'){\cal A}_1^T \label{eq.:cal-B} \\
{\cal R}'&\triangleq \diag\Big \{ \sum_{\ell \in {\cal N}_k}c_{\ell,k} \big(R_{u,\ell}+\sigma^2_{n,\ell}I_M\big), k=1,\cdots,N \Big\}\\
{\cal P}'&\triangleq\diag\Big \{ \sum_{\ell \in {\cal N}_k}c_{\ell,k}\, \sigma^2_{n,\ell}I_M,\; k=1,\cdots,N \Big\}.
\end{align}
As it is clear from (\ref{eq.:mean_perfomance-biased}), the bias is created by the regression noise $\{\n_{k,i}\}$ only, whereas the noise $\{\v_k(i)\}$ has no effect on generating the bias.
\end{Remark}
\subsection{Mean-Square Convergence and Stability}
\label{subsec.:MeanSquareStability}
To study the mean-square performance of the proposed algorithms, we first follow the energy conservation arguments of \cite{cattivelli2010diffusion,sayed2008} and determine a variance relation that is suitable in the current context. The relation can be obtained in the limit, as $i\rightarrow\infty$, by computing the expectation of the weighted squared norm of (\ref{eq.:global_error_vector_w}) under Assumption \ref{assump.:data model}:
\begin{align}
\E\|\tilde \w_i\|^2_{\Sigma}=&\E\Big(\|\tilde \w_{i-1}\|^2_{{\boldsymbol{\Sigma}}'}\Big)+\E[\boldsymbol{{g}}^{\ast}_i {\mathcal M}{\cal A}_2\Sigma{\cal A}_2^T{\mathcal M}\boldsymbol{g}_i] \nonumber \\
&+\E[{ \omega^o}^{\ast} \boldsymbol{\mathcal {P}}_i^{\ast}{\mathcal M}{\cal A}_2\Sigma {\cal A}_2^T{\mathcal M}\boldsymbol{\mathcal {P}}_i \omega^o]
\label{variance_relation_1}
\end{align}
where $\|x\|^2_{\Sigma}=x^* \Sigma x$  and $\Sigma \geq 0$ is a weighting matrix that we are free to choose. Note that (\ref{variance_relation_1}) is obtained by eliminating the following terms:
\begin{align}
&\E[({\cal A}_2^T{\mathcal M} \boldsymbol{g}_i)^{\ast}\Sigma{\cal A}_2^T(I-{\mathcal M}\boldsymbol{\mathcal{R}}_i){\mathcal A_1}^T \Sigma \tilde \w_{i-1}]=0 \\
&\E[({\cal A}_2^T(I-{\mathcal M}\boldsymbol{\mathcal{R}}_i){\mathcal A_1}^T \tilde \w_{i-1})^{\ast}\Sigma {\cal A}_2^T{\mathcal M} \boldsymbol{g}_i]=0 \\
&\E[({\cal A}_2^T{\mathcal M}\boldsymbol{\mathcal {P}}_i w^o)^{\ast}\Sigma{\cal A}_2^T(I-{\mathcal M}\boldsymbol{\mathcal{R}}_i){\mathcal A_1}^T \tilde \w_{i-1}]=0 \\
&\E[({\cal A}_2^T(I-{\mathcal M}\boldsymbol{\mathcal{R}}_i){\mathcal A_1}^T \tilde \w_{i-1})^{\ast}\Sigma  {\cal A}_2^T{\mathcal M}\boldsymbol{\mathcal {P}}_i w^o]=0.
\end{align}
These terms are zero firstly, because $\tilde \w_{i-1}$ is independent of $\boldsymbol{g}_i$, $\boldsymbol{\mathcal {P}}_i$ and $\boldsymbol{\mathcal{R}}_i$ under Assumption \ref{assump.:data model} \cite{Zhao2012imperfect}, and secondly, since the proposed algorithms are unbiased, $\E[\tilde \w_i]$ is zero for large $i$, if the step-sizes are chosen as in (\ref{eq.:mu_range}). 

In relation (\ref{variance_relation_1}), we have:
\begin{align}
{\boldsymbol{\Sigma}}'={\boldsymbol{{\cal B}}_i}^{\ast} \Sigma \boldsymbol{{\cal B}}_i.
\end{align}
It follows from Assumption \ref{assump.:data model} that $\tilde{\w}_{i-1}$ and $\boldsymbol{\cal R}_i$ are independent of each other so that
\begin{align}
\E \Big(\|\tilde \w_{i-1}\|^2_{{\boldsymbol{\Sigma}}'}\Big)=\E\|\tilde \w_{i-1}\|^2_{\E[{\boldsymbol{\Sigma}}']}.
\end{align}
Substituting this expression into (\ref{variance_relation_1}), we arrive at:
\begin{align}
\E\|\tilde \w_i\|^2_{\Sigma}&=\E\|\tilde \w_{i-1}\|^2_{{\Sigma}'}+\Tr[\Sigma {\cal A}_2^T {\mathcal M}{\mathcal
G}{\mathcal M}{\cal A}_2]\nonumber \\
&\hspace{1cm}+\Tr[\Sigma {\cal A}_2^T {\mathcal M}\Pi{\mathcal M}{\cal A}_2]
\label{variance_relation_2}
\end{align}
where
\be
{\Sigma}'=\E[\boldsymbol{{\cal B}^*_i}\Sigma \boldsymbol{\mathcal{B}}_i].
\label{eq:SigmaPrime}
\ee
In equation (\ref{variance_relation_2}) ${\mathcal G}=\E[\boldsymbol{g}_i\boldsymbol{g}^{\ast}_i]$, which using (\ref{eq.:Gi-deginition}) is given by (see Appendix \ref{apex.:mathcalG_calculation}):

\small
\begin{align}
{\mathcal G}={\cal C}^T \diag \Big \{\sigma^2_{v,1}(R_{u,1}+\sigma^2_{n,1}I),\ldots,\sigma^2_{v,N}(R_{u,N}+\sigma^2_{n,N}I) \Big \}{\cal C}.
\label{eq.:def_G}
\end{align}
\normalsize
In relation (\ref{variance_relation_2}), ${\Pi}=\E[\boldsymbol{\mathcal {P}}_i \omega^o{ \omega^o}^{\ast}\boldsymbol{\mathcal {P}}^{\ast}_i]$ and its $(k,j)$-th block is computed as (see Appendix \ref{apex.:mathcalN_calculation}):
\begin{align}
\Pi_{k,j}=&\sum_{\ell} c_{\ell,k}c_{\ell,j}\Big\{\sigma^2_{n,\ell} \|w^o\|^2 \big(R_{u,\ell}+ \sigma^2_{n,\ell} I\big)\nonumber \\
&\hspace{3cm}+(\beta-1)\sigma^4_{n,\ell}w^o w^{o*}\Big\}
\end{align}
where $\beta=2$ for real-valued data and $\beta=1$ for complex-valued data. 
If we introduce $\sigma=\bvec(\Sigma)$ and ${\sigma}'=\bvec({\Sigma}')$ then we can write $\sigma'={\cal F}\sigma$ where
\be
{\cal F}=\E[\boldsymbol{{\cal B}^T_i}\otimes_b \boldsymbol{\mathcal{B}}^*_i]
\label{eq:exact_calF}
\ee
Considering these definitions, the variance relation in (\ref{variance_relation_2}) can be rewritten more compactly as:
\begin{align}
\E\|\tilde \w_i\|^2_{\sigma}=&\E\|\tilde \w_{i-1}\|^2_{{\cal F}\sigma}+\gamma^T \sigma
\label{variance_relation_3}
\end{align}
where we are using the notation  $\|x\|^2_{\sigma}$ as a short form for $\|x\|^2_{\Sigma}$, and where
\begin{align}
\gamma=\bvec({\cal A}_2^T {\mathcal M}{\mathcal G}^T{\mathcal M}{\mathcal
A_2}+{\cal A}_2^T {\mathcal M} \Pi^T{\mathcal M}{\cal A}_2).
\end{align}
To compute ${\cal F}$, we expand ${\Sigma}'$ from
(\ref{eq:SigmaPrime}) to get:
\begin{align}
{\Sigma}'=&{\mathcal A_1}\Big({\cal A}_2\Sigma{\cal A}_2^T-{\mathcal R}{\mathcal M}{\cal A}_2\Sigma{\cal A}_2^T- {\cal A}_2\Sigma{\cal A}_2^T{\mathcal M}{\mathcal R}\Big ){\cal A}_1^T \nonumber \\
&+\E[{\mathcal A_1}\boldsymbol{\mathcal{R}}^{\ast}_i{\mathcal M}{\cal A}_2\Sigma{\cal A}_2^T{\mathcal M}\boldsymbol{\mathcal{R}}_i{\cal A}_1^T].
\label{eq.:sigma_prime_value}
\end{align}
The last term in (\ref{eq.:sigma_prime_value}) depends on ${\mathcal M}^2$ and can, therefore, be neglected for small step-sizes.
As a result, we obtain
\small
\begin{align}
{\cal F}\approx({\mathcal A_1}\otimes_{\scrsmall{b}}{\mathcal A_1})(I-I\otimes_{\scrsmall{b}}{\mathcal R}{\mathcal M}-{\mathcal R}^T{\mathcal M}\otimes_{\scrsmall{b}} I)({\cal A}_2\otimes_{\scrsmall{b}} {\mathcal
A_2}).
\label{eq.:small_stepsize_F}
\end{align}
\normalsize
We can also derive a more compact expression to compute $\cal F$. To this end, we first note that the last term in (\ref{eq.:sigma_prime_value}) can be expressed as:
\begin{align}
\E[{\mathcal A_1}\boldsymbol{\mathcal{R}}^{\ast}_i{\mathcal M}{\cal A}_2\Sigma{\cal A}_2^T{\mathcal M}\boldsymbol{\mathcal{R}}_i{\cal A}_1^T]&=\E[{\mathcal A_1}{\mathcal{R}}^{\ast}{\mathcal M}{\cal A}_2\Sigma{\cal A}_2^T{\mathcal M}{\mathcal{R}}{\cal A}_1^T]\nonumber \\
&\hspace{1cm}+O({\cal M}^2)
\label{eq:last-term-approx}
\end{align}
Now by substituting (\ref{eq:last-term-approx}) into (\ref{eq.:sigma_prime_value}) and ignoring the remaining terms that depend on ${\cal M}^2$, under the small step-size condition, we arrive at:
\begin{align}
{\cal F}&\approx {{\cal B}}^T\otimes_{\scrsmall{b}} {{\cal B}}^*
\label{eq.:ApproximateF}
\end{align} 

We now proceed to show the stability of the algorithm in the mean-square error sense, as follows.
Using (\ref{variance_relation_3}), we can write:
\begin{equation}
\lim_{i\rightarrow \infty}\E\|\tilde \w_{i}\|^2_{\sigma}=\lim_{i\rightarrow \infty}\E\|\tilde \w_{-1}\|^2_{{\cal F}^{i+1}{\sigma}}+ \gamma^T\sum_{j=0}^{\infty} {\cal F}^j\sigma.
\label{eq.:transient_recursion-stability}
\end{equation}
As it is evident from this expression, the proposed algorithms will be stable in the mean-square sense if ${\cal F}$ is stable.
From (\ref{eq.:ApproximateF}), we deduce that ${\cal F}$ will be stable if ${\cal B}$ is stable. According to our mean-convergence analysis, the stability of ${\cal B}$ is guaranteed if (\ref{eq.:mu_range}) holds. Therefore, the step-size condition (\ref{eq.:mu_range}) is sufficient to guarantee the stability of the algorithms both in the mean and mean-square sense.

\subsection{Mean-Square Steady-State Performance}
\label{subsec.:SteadyStateMeanSquareAnalysis}
To obtain mean-square error (MSE) steady state expressions for the network, we let $i$ go to infinity and use expression (\ref{variance_relation_3}) to write:
\begin{align}
\lim_{i \rightarrow \infty}\E\|\tilde \w_i\|^2_{(I-{\cal F})\sigma}=\gamma^T\sigma.
\label{eq.tild-wi-infinity}
\end{align}
By definition, the MSD and EMSE at each node $k$ are respectively computed as:
\begin{align}
\eta_k=\lim_{i \rightarrow \infty}\E\|\tilde \w_{k,i}\|^2,\quad \zeta_k=\lim_{i \rightarrow \infty}\E\|\tilde \w_{k,i}\|^2_{R_{u,k}}.
\end{align}
The MSD and EMSE of the nodes can be retrieved from the network error vector $\tilde \w_i$  by writing:
\begin{align}
\eta_k&=\lim_{i \rightarrow \infty}\E\|\tilde \w_i\|^2_{\{\diag(e_k) \otimes I\}} \label{eq.:msd-k-definition}\\
\zeta_k&=\lim_{i \rightarrow \infty}\E\|\tilde \w_i\|^2_{\{\diag( e_k)\otimes R_{u,k}\}} \label{eq.emse-k-definition}
\end{align}
where $e_k$ is a canonical basis vector in ${\amsbb R}^N$ with entry one at position $k$. From (\ref{eq.tild-wi-infinity}) and (\ref{eq.:msd-k-definition}), we can obtain the MSD at node $k$, for $k\in\{1,2,\cdots,N\}$:
\begin{align}
\eta_k=\gamma^T (I-{\cal F})^{-1}\bvec(\diag(e_k)\otimes I_M).
\end{align}
In the same manner, we compute the EMSE at node $k$ as:
\begin{align}
\zeta_k=\gamma^T (I-{\cal F})^{-1}\bvec\big(\diag(e_k)\otimes R_{u,k}\big).
\end{align}
The network MSD and EMSE are defined as the average of MSD and EMSE values over the network, i.e.,
\begin{align}
&\eta=\frac{1}{N}\sum_{k=1}^N \eta_k, \quad \zeta=\frac{1}{N}\sum_{k=1}^N \zeta_k.
\label{eq.:net_emse_msd_steady_state}
\end{align}

\subsection{Mean-Square Transient Behavior}
We use (\ref{variance_relation_3}) to obtain an expression for the mean-square behavior of the algorithm in transient-state.
In this expression, if we substitute  $ \w_{k,-1}=0,\; \forall k \in\{1,\cdots,N\}$, we obtain:
\begin{equation}
\|\tilde \w_{i}\|^2_{\sigma}=\| w^o\|^2_{{\cal F}^{i+1}{\sigma}}+ \gamma^T\sum_{j=0}^i {\cal F}^j\sigma.
\label{eq.:transient_recursion2}
\end{equation}
Writing this recursion for $i-1$, and subtract it from (\ref{eq.:transient_recursion2}) leads to:
\begin{equation}
\|\tilde \w_{i}\|^2_{\sigma}=\|\tilde \w_{i-1}\|^2_{\sigma}+\| w^o\|^2_{{\cal F}^i(I-{\cal F}){\sigma}}+ \gamma^T {\cal F}^i\sigma.
\label{eq.:transient_recursion3}
\end{equation}
By replacing $\sigma$  with $\sigma_{\msd_k}=\bvec\big(\diag\{e_k\}\otimes I_M)$ and $\sigma_{\emse_k}=\bvec\big(\diag\{e_k\}\otimes R_{u,k})$ and using $\w_{k,-1}=0$, we arrive at the following two recursions for the evolution of MSD and EMSE over time:

\small
\begin{align}
&\eta_k(i)=\eta_k(i-1)-\|w^o\|_{{\cal F}^i(I-{\cal F}){\sigma}_{\msd_k}}+\gamma^T {\cal F}^i \sigma_{\msd_k} \label{eq.:msd-transient_state_noisy_regressor}\\
&\zeta_k(i)=\zeta_k(i-1)-\|w^o\|_{{\cal F}^{i-1}(I-{\cal F}){\sigma}_{\emse_k}}+\gamma^T {\cal F}^{i-1} \sigma_{\emse_k}.
\label{eq.:emse-transient_state_noisy_regressor}
\end{align}
\normalsize
The MSD and EMSE of the network can be computed either by averaging the nodes transient behavior, or by substituting
\begin{align}
&\sigma_{\msd}=\frac{1}{N} \bvec(I_{MN}) \label{eq.:sigma-msd-network}\\
&\sigma_{\emse}=\frac{1}{N} \bvec\big(\diag\{R_{u,1},\cdots,R_{u,N}\}\big)
\label{eq.:sigma-emse-network}
\end{align}
in recursion (\ref{eq.:transient_recursion3}).  We summarize the mean-square analysis results of the algorithms in the following:
\begin{thm}
\label{theo:DiffLmsErroWectorknownVariance}
Consider an adaptive network operating under bias-compensated diffusion  Algorithm \ref{alg.:ATC Bias-Compensated Diffusion LMS} or \ref{alg.:CTA Bias-Compensated Diffusion LMS} with the space-time data (\ref{eq.:network_linear_model_eq1}) and (\ref{eq.:network_linear_model_eq2}) that satisfy Assumption \ref{assump.:data model}. In this network, if we assume that the regressors noise variances are known or perfectly estimated and nodes initialize at zero, then the MSD and EMSE of each node $k$ evolve with time according to (\ref{eq.:msd-transient_state_noisy_regressor}) and (\ref{eq.:emse-transient_state_noisy_regressor}) and the network MSD and EMSE follow recursions:
\begin{align}
&\eta(i)={\eta}(i-1)-\|w^o\|_{{\cal F}^i(I-{\cal F}){\sigma}_{\msd}}+\gamma^T {\cal F}^i \sigma_{\msd} \nonumber\\
&\zeta(i)={\zeta}(i-1)-\|w^o\|_{{\cal F}^{i-1}(I-{\cal F}){\sigma}_{\emse}}+\gamma^T {\cal F}^{i-1} \sigma_{\emse} \nonumber
\label{eq.:transient_state_noisy_regressor}
\end{align}
where ${\sigma}_{\msd}$, and ${\sigma}_{\emse}$ are defined in (\ref{eq.:sigma-msd-network}) and (\ref{eq.:sigma-emse-network}) and $\cal F$ is given by (\ref{eq:exact_calF}). Moreover, if the step-sizes are chosen to satisfy (\ref{eq.:mu_range}), the network will be stable, converge in the mean and mean-square sense and reach the steady-state MSD and EMSE characterized by (\ref{eq.:net_emse_msd_steady_state}).
\end{thm}

%
\section{Regression Noise Variance Estimation}
\label{subsec.:RegressorNoiseVarianceEstimation}
In the proposed algorithms, each node still needs to have the regression noise variances, $\{\sigma^2_{n,\ell}\}_{k=1}^{\mathcal{N}_k}$,  to evaluate the stochastic gradient vector, $\widehat{\nabla  J_{\ell}}$. In practice, such information is rarely available and normally obtained through estimation.
A review of previous works reveals that the regression noise variances can be either estimated off-line \cite{bertranddiffusion2011}, or in real-time when the unknown parameter vector, $w^o$, is being estimated \cite{zheng2003least,jia2009forward}.
For example, in the context of speech analysis, they can be estimated off-line during silent periods in between words and sentences \cite{bertranddiffusion2011}. In some other applications, these variances are estimated during the operation of the algorithm using the second-order moments of the regression data and the system output signal \cite{zheng2003least,jia2009forward}. In what follows we propose an adaptive recursive approach to estimate the regression noise variances without using the second order moments of the data.

The variance of the regression noise at each node is classified as local information and, hence, it can be estimated from the node's local data. When the regression data at node $k$ is not corrupted by measurement noise (i.e., $\z_{k,i}=\u_{k,i}$), and when the node operates independent of all other nodes to estimate $w^o$ by minimizing $\E|\d_k(i)-\u_{k,i} w|^2$, the minimum attainable MSE can be expressed as \cite{sayed2008}:
\be
J_{\rm min} \triangleq \sigma^2_{d,k}-r^*_{du,k}R_{u,k}^{-1}r_{du,k}.
\label{eq.:j_min_unbiased0}
\ee
Under noisy regression scenarios where node $k$ operates independently to minimize the cost (\ref{eq.:modified-cost-node-k}), the minimum achievable cost will still be (\ref{eq.:j_min_unbiased0}).
To verify this, we note from Remark \ref{re:Jk-positive-definite} that since $J_{k}(w)$ is positive definite and, hence, strongly convex, its unique minimizer under Assumption \ref{assump.:data model} will be $w^o$. Therefore, substituting $w^o$ into (\ref{eq.:modified-cost-node-k}) will give its minimum, i.e.:
\begin{align}
\min_w J_{k}(w) &=\E| \d_k(i)- \z_{k,i} w^o|^2-\sigma^2_{n,k}\| w^o\|^2 \nonumber \\
&=\sigma^2_{d,k}-r^*_{du,k} R_{u,k}^{-1}r_{du,k} \label{eq.:j_min_unbiased3} \\
&=J_{\rm min}. \label{eq.:j_min_unbiased3}
\end{align}
We use this result to estimate the regression noise variance $\sigma^2_{n,k}$ at each node $k$.

Now, let us introduce
\be
\e_k(i) \triangleq \d_k(i)-\z_{k,i} \w_{k,i-1}
\ee
where $\w_{k,i-1}$ is the weight estimate from ATC diffusion (which would be replaced by $\bpsi_{k,i-1}$ for CTA diffusion). Considering $J_{k}(\w_{k,i-1})$, for sufficiently small step-sizes and in the limit when the weight estimate is close enough to $w^o$, it holds that:
\be
\E |\e_k(i)|^2-\sigma^2_{n,k} \| w^o\|^2\approx J_{\rm min}.
\label{eq.lim-e2k}
\ee
From (\ref{eq.:network_linear_model_eq2}) and (\ref{eq.:j_min_unbiased0}), it can be verified that $J_{\rm min}=\sigma^2_{v,k}$, and hence from (\ref{eq.lim-e2k}), we can write:
\be
 \E |\e_k(i)|^2\approx \sigma^2_{v,k}+\sigma^2_{n,k} \| w^o\|^2.
\label{eq.lim-e2k-2}
\ee
In this relation, $\sigma^2_{v,k}$, can be ignored if $\sigma^2_{n,k} \| w^o\|^2 \gg \sigma^2_{v,k}$.
Under such circumstances, if we assume $\|w^o\|^2\neq 0$, which is true for systems with at least one non-zero coefficient, then the variance of the regression noise can be obtained by:
\vspace{-0.1cm}
\begin{align}
\sigma^2_{n,k}\approx\frac{\E |\e_k(i)|^2}{\|w^o\|^2}.
\label{eq.:sigma2n_approximation1}
\end{align}
Since, in (\ref{eq.:sigma2n_approximation1}), $\E |\e_k(i)|^2$ and the unknown parameter, $w^o$, are initially unavailable, we can estimate $\sigma^2_{n,k}$ using the following relations as the latest estimates of these quantities become available, i.e.,
\vspace{-0.1cm}
\begin{align}
\f_{k}(i)&=\alpha \f_{k}(i-1)+(1-\alpha)|\e_k(i)|^2
\label{eq.:RegressorNoiseVarianceComputation1} \\
{\bsigma}^2_{n,k}(i)&=\frac{\f_k(i)}{\|\w_{k,i}\|^2}
\label{eq.:RegressorNoiseVarianceComputation2}
\end{align}
where $0\ll \alpha < 1$ is a smoothing factor with nominal values in the range of $[0.95, 0.99]$.
\begin{assump}
\label{assump.:RegressionNoiseVarianceEst}
The regression noise variance, $\sigma^2_{n,k}$,  and the output measurement noise, $\sigma^2_{v,k}$, satisfy the following inequality
\be
\sigma^2_{n,k} \| w^o\|^2 \gg \sigma^2_{v,k}.
\label{eq.:model-noise-condition}
\ee
\end{assump}
Under this assumption, the regressor noise variance at each node $k$ can be adaptively estimated via (\ref{eq.:RegressorNoiseVarianceComputation1}) and (\ref{eq.:RegressorNoiseVarianceComputation2}) using the data samples $\e_k(i)$  and $\w_{k,i-1}$ supplied from the bias-compensated LMS iterations.

\section{Simulation Results}
\label{sec.:UnbiasedDiffLmsResults}
\begin{figure}
\centering
\includegraphics[width=0.8 \columnwidth]{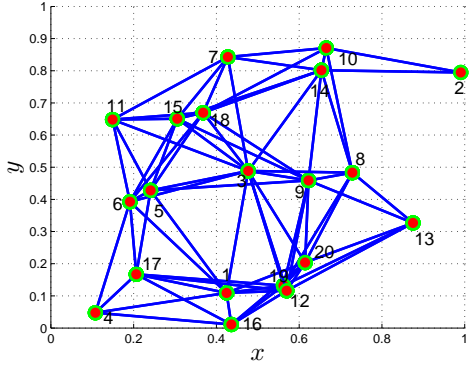}
\caption{\small{Network topology used in the simulations.}}
\label{fig.:network_topology}
\end{figure}
In this section, we present computer experiments to illustrate the efficiency of the proposed algorithms and to verify the theoretical findings.
We evaluate the algorithm performance for known regressor noise variance and with adaptive noise variance estimation. We consider a connected network with $N=20$ nodes that are positioned randomly on a unit square area with maximum communication distance of $0.4$ unit length. The network topology is shown in Fig. \ref{fig.:network_topology}. We choose $A_1=I$, compute $A_2$ using the relative-variance rule (\ref{eq.:relative_variance}) and choose the matrix $C$ according to the metropolis criterion \cite{lopes2008diffusion,sayed2012diffusion}. In the plots, we use $A_{\text{rel}}$ and $C_{\text{met}}$ to refer to this particular choice of $A_2$ and $C$. The network data are generated according to model (\ref{eq.:network_linear_model_eq1}) and (\ref{eq.:network_linear_model_eq2}). The aim is to estimate the system parameter vector $w^o={[1, 1]^T}/{\sqrt{2}}$ over the network using the proposed bias-compensated diffusion algorithms. In all our experiments, the curves from the simulation results are drawn from the average of $500$ independent runs.

We choose the step-sizes as $\mu_k=0.05$, and set $\w_{k,-1}=[0,0]^T$, for all $k$.
We adopt Gaussian distribution to generate $\v_{k}(i)$, $\n_{k,i}$ and $\u_{k,i}$. The covariance matrices of the regression data and the regression noise are of the form $R_{u,k}=\sigma^2_{u,k} I_M$, and $\sigma^2_{n,k} I_M$, respectively. The network signal and noise power profile, are given in Table \ref{ta:.noise_info}.

\begin{figure}
\centering
\includegraphics[width=\columnwidth]{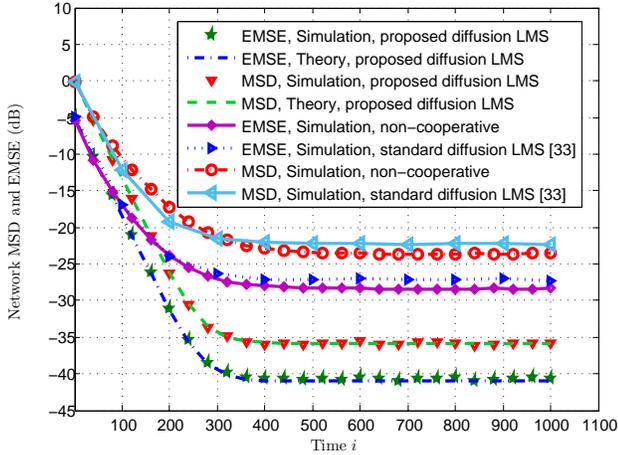}
\caption{\small{Convergence behavior of the proposed bias-compensated diffusion LMS, standard diffusion LMS and non-cooperative LMS algorithms.}}
\label{fig.:network_transient_analysis}
\end{figure}

\begin{table}
\caption{\small{Network signal and noise power profile}}
\begin{center}
\small
\begin{tabular}{ | c | c | c | c |  }
     \hline
                            & \multicolumn{3}{|c|}{\bf{Parameters}}        \\ \hline
      \bf{Node $k$}         & $\sigma^2_{v,k}$       &$\Tr(R_{u,k})$     &$\sigma^2_{n,k}$              \\ \hline
       1                    &   0.0230                &  0.3000                                         &  0.0170                  \\ \hline
       2                    & 0.0020                 &  0.7500                                         &  0.0970                    \\ \hline
       3                    &  0.0160                & 0.5250                                           &  0.0620                   \\ \hline
       4                    &  0.0040               & 0.4250                                         &    0.0570                    \\ \hline
       5                    & 0.0420                 & 0.6000                                           &  0.0600                 \\ \hline
       6                    & 0.0400                & 0.6500                                         & 0.0730                  \\ \hline
       7                    & 0.0120               & 1.0000                                         &  0.0560                  \\ \hline
       8                    & 0.0120                 & 0.7750                                          &  0.0860                   \\ \hline
       9                    &  0.0310               & 0.7250                                          &   0.0250               \\ \hline
       10                   & 0.0280                & 0.6750                                           &  0.0490                 \\ \hline
       11                   &  0.0350                & 0.6500                                         &  0.0680                 \\ \hline
       12                   &  0.0500                & 0.6000                                         &0.0760                  \\ \hline
       13                   &   0.0090               & 0.2750                                          &0.0600                  \\ \hline
       14                   &  0.0340              & 0.3500                                         & 0.0150                 \\ \hline
       15                   & 0.0290               & 0.6250                                        & 0.0160                 \\ \hline
       16                   & 0.0280                & 0.9250                                      &0.0490                 \\ \hline
       17                   & 0.0020                &0.3250                                          & 0.0830                 \\ \hline
       18                   &  0.0080                 &0.8750                                          &0.0370                  \\ \hline
       19                   & 0.0410               &0.2500                                            &0.0170                  \\ \hline
       20                   &  0.0460               &0.8000                                          & 0.0160                 \\ \hline

\end{tabular}
\end{center}
\label{ta:.noise_info}
\end{table}
\paragraph{Transient MSE Results with Perfect Noise Variance Estimation}
In Fig. \ref{fig.:network_transient_analysis}, we demonstrate the network transient behavior in terms of MSD and EMSE for the proposed diffusion LMS algorithm, standard diffusion LMS algorithm \cite{cattivelli2010diffusion} and the non-cooperative mode of the proposed algorithm. Note that $A_2=I$ and $C=I$ correspond to the non-cooperative network mode of the proposed algorithm, where each node runs a stand alone bias-compensated LMS. As the results indicate, the performance of the cooperative network with $C_{\text{met}}$ and $A_{\text{rel}}$ exceeds that of the non-cooperative case by 12 dB.  We also observe that the proposed algorithm outperform the standard diffusion LMS \cite{cattivelli2010diffusion} by more that 12dB. It is interesting to note that the non-cooperative algorithm outperforms the standard diffusion LMS by about 1dB.

\begin{figure}
\centering
\includegraphics[width=\columnwidth]{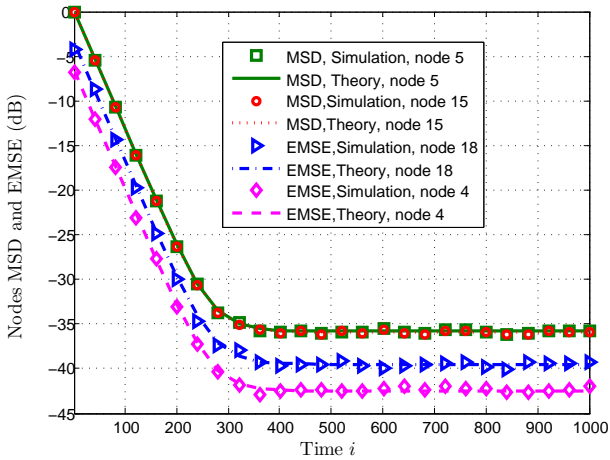}
\caption{\small{ MSD learning curves of nodes $5$ and $15$ and EMSE learning curves of nodes $4$ and $18$.}}
\label{fig.:node_trans}
\end{figure}
\begin{figure}
\centering
\includegraphics[width=\columnwidth]{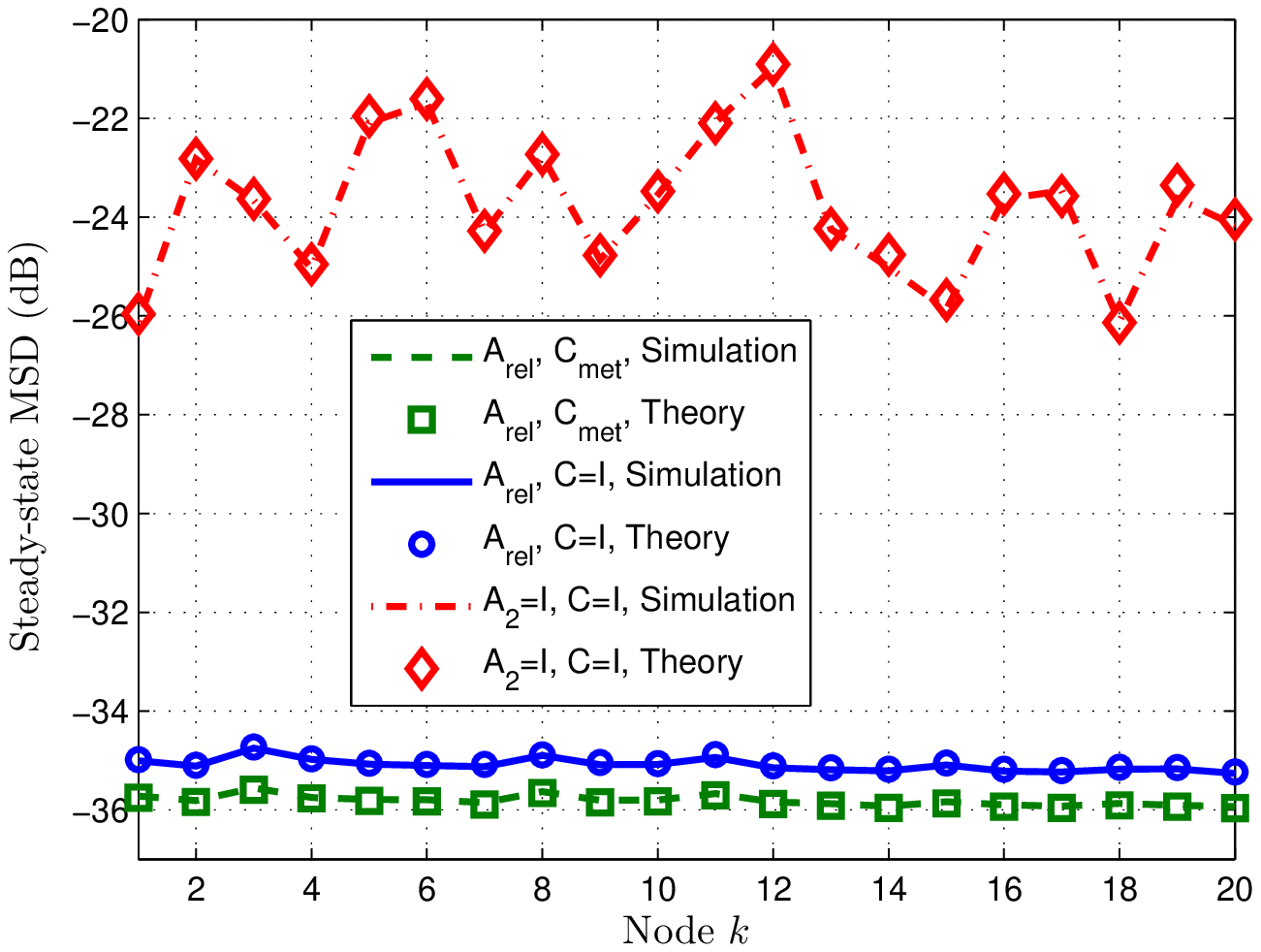}
\caption{\small{Network steady-state MSD for different combination matrices.}}
\label{fig.:steady_state_msd}
\end{figure}
\begin{figure}
\centering
\includegraphics[width=\columnwidth]{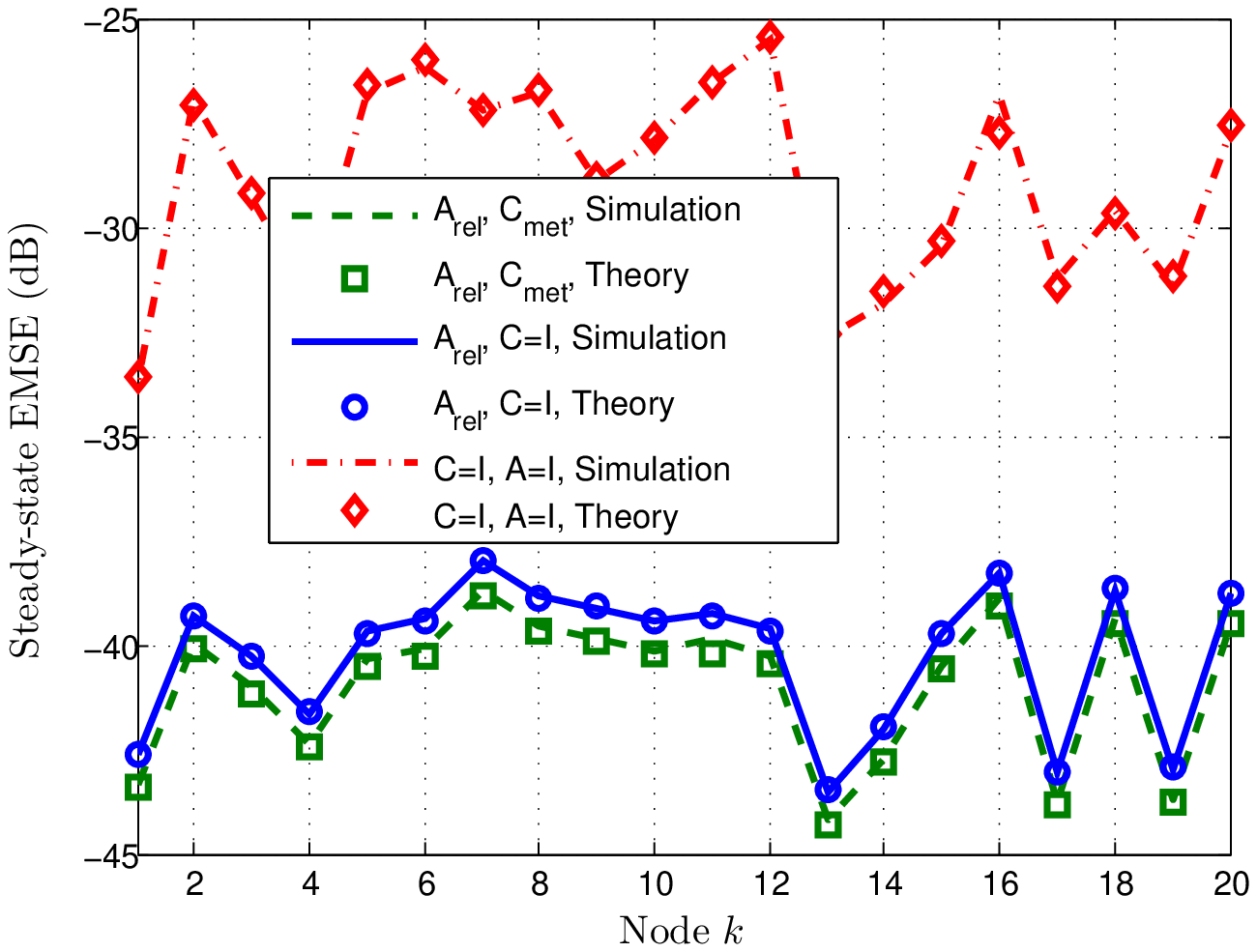}
\caption{\small{Network steady-state EMSE for different combination matrices.}}
\label{fig.:steady_state_emse}
\end{figure}
\begin{figure}
\centering
\includegraphics[width=\columnwidth]{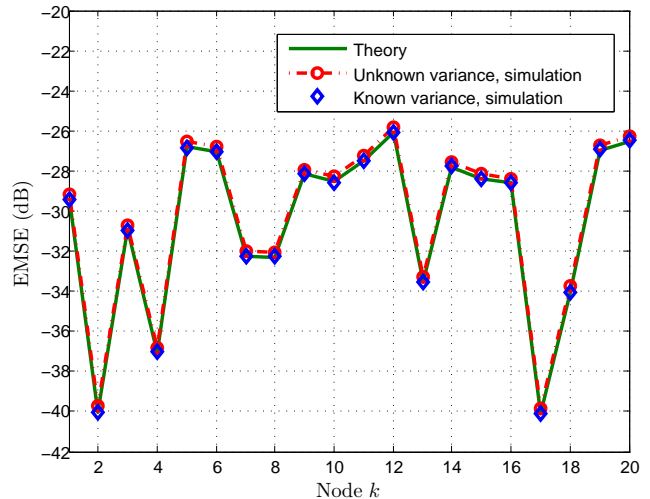}
\caption{\small{Steady-state network EMSE with known and estimated regressor noise variances.}}
\label{fig.:steady_state_emse_unknown_variance}
\end{figure}
We also present the EMSE and MSD of some randomly chosen nodes  in Fig. \ref{fig.:node_trans}. In particular, we plot the EMSE learning curves of nodes $4$ and $18$ and the MSD learning curves of nodes $5$ and $15$. We observe that the MSD curves of the chosen nodes are identical. Since the algorithm is unbiased, this implies that these nodes have reached agreement about the unknown network parameter, $w^o$. As we will show in the steady-state results, all nodes over the network almost reach agreement.  We note that, in all scenarios, there is a good agreement between simulations and the analysis results.
%

\paragraph{Steady-State MSE Results with Perfect Noise Variance Estimation}
The network steady-state MSD and EMSE  are shown in Figs. \ref{fig.:steady_state_msd} and \ref{fig.:steady_state_emse}. From these figures, we observe that there is a good agreement between simulations and analytical findings. In addition, we consider the case when nodes only exchange their intermediate estimates (i.e., when $C=I$). It is seen that the  MSD performance of the algorithm with $C_{\text{met}}$ is $1$dB superior than that with $C=I$. We also observe that the performance discrepancies between nodes in terms of MSD is less than $0.5$dB for cooperative scenarios, while in the non-cooperative scenario it is more than $5$dB. This shows agreement in the network in spite of different noise and energy profiles at each node. Note that the fluctuations in EMSE over the network are due to differences in energy level in the nodes' input signals, but this does not preclude the cooperating nodes from reaching a consensus in the estimated parameters.
\begin{figure}
\centering
\includegraphics[width=\columnwidth]{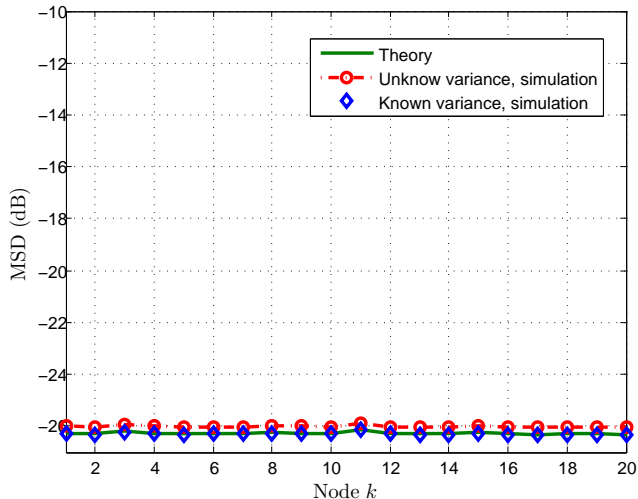}
\caption{\small{Steady-state network MSD with known and estimated regressor noise variances.}}
\label{fig.:steady_state_msd_unknown_variance}
\end{figure}
\begin{figure}
\centering
\includegraphics[width=\columnwidth]{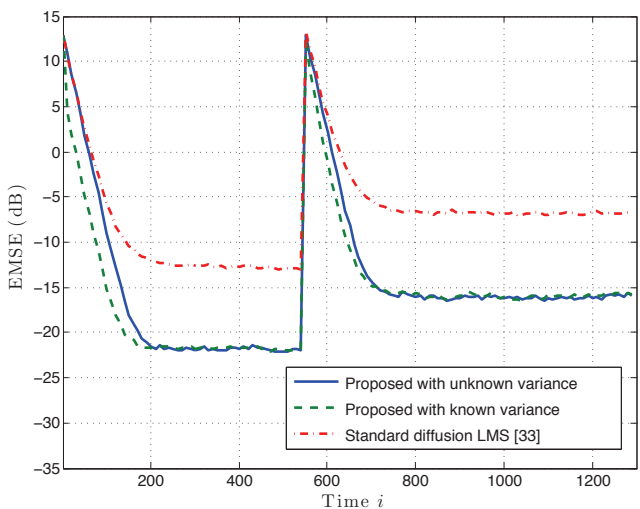}
\caption{\small{EMSE Tracking performance with known and estimated regressor noise variances.}}
\label{fig.:EMSE_tracking_performance}
\end{figure}
\begin{figure}
\centering
\includegraphics[width=\columnwidth]{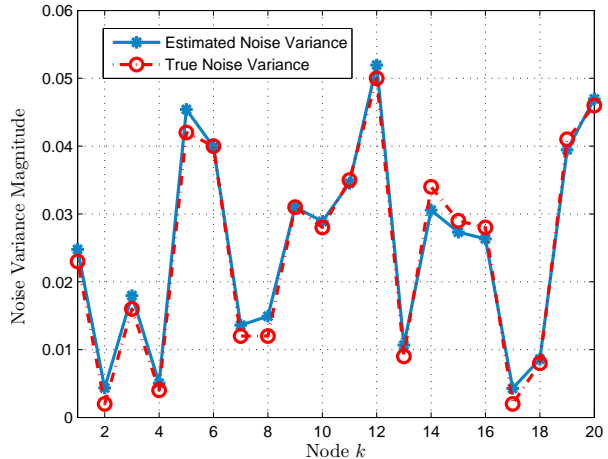}
\caption{\small{The estimated and true value of the regression noise variance, $\sigma^2_{n,k},$ over the network.}}
\label{fig.:regressionNoiseVarianceEstimate}
\end{figure}

\paragraph{MSE Results of the Algorithm with Adaptive Noise Variance Estimation}
We compare the transient and steady-state behavior of the bias-compensated diffusion LMS with known regressor noise variance and adaptive noise variance estimation. For this experiment, we consider the same network topology and noise profile as above. However, the unknown parameter vector to be estimated, in this case, is $w^o=2{\mathbb 1}_5+2j{\mathbb 1}_5$, where ${\mathbb 1}_M$ is a $M\times1$ column vector with unit entries. The network energy profile is chosen as $\Tr(R_{u,k})=20 \Tr(\sigma^2_{n,k}I)$. Using these choices, Assumption \ref{assump.:RegressionNoiseVarianceEst}  will be satisfied. We set $\alpha=0.99$ and $\mu_k=0.01$ for all $k$.

Figs. \ref{fig.:steady_state_emse_unknown_variance} and  \ref{fig.:steady_state_msd_unknown_variance}  show the steady-state EMSE and MSD of the network for these two cases. 
The steady-state values are obtained by averaging over the last 200 samples after initial convergence. We observe that the performance of the proposed bias-compensated LMS algorithm with adaptive noise variance estimation is almost identical to that of the ideal case with known noise variances.

Fig. \ref{fig.:EMSE_tracking_performance} illustrates the tracking performance of the bias-compensated diffusion LMS algorithm for these two cases for a sudden change in the unknown parameter $w^o$ and compares the results with that of the standard diffusion LMS algorithm given in \cite{cattivelli2010diffusion}. The variation in the unknown parameter vector occurs at iteration $i=550$ when $w^o$ changes to $2\,w^o$. Similar conclusion as in Fig. \ref{fig.:steady_state_emse_unknown_variance} and  \ref{fig.:steady_state_msd_unknown_variance}  can be made for the proposed algorithms with known and estimated regression noise variances. We also observe that the proposed algorithms outperform the standard diffusion LMS \cite{cattivelli2010diffusion} by nearly 10dB in steady-state.

Fig.~\ref{fig.:regressionNoiseVarianceEstimate} illustrates the results of regression noise variance estimation in the steady state. In this experiment, we observe that for $i \ge 350$,  $\E[\bsigma^2_{n,k}(i)] \rightarrow \sigma^2_{n,k}$. This indicates that the proposed adaptive estimation strategy  for computation of the nodes' regression noise variance over the network works well.

\section{Conclusion}
\label{sec.:UnbiasedDiffLMSConclusion}
We developed bias-compensated diffusion LMS strategies for parameter estimation over sensor networks where the regression data are corrupted with additive noise. The algorithms operate in a distributed manner and exchange data via single-hop communication to save energy and communication resources. The proposed algorithms estimate the regression noise variances and use them to remove the bias from the estimate. In the analysis, it has been shown that the proposed bias-compensated diffusion algorithms are unbiased and converge in the mean and mean-square error sense for sufficiently small step-sizes.
We carried out computer experiments that confirmed the effectiveness of the algorithms and support the analytical findings.

\begin{appendices}  
\section{computation of ${\mathcal G}$}
\label{apex.:mathcalG_calculation}
This can be computed by substituting $\boldsymbol{g}(i)$ from (\ref{eq.:Gi-deginition}) into ${\mathcal G}= \E[\boldsymbol{g}_i\boldsymbol{g}^{\ast}_i]$ , as a result:
\begin{align}
 {\mathcal G}=&{\mathcal C}^T \E
\left [\begin{array}{c}
   \z_{1,i}^{\ast} \v_{1}(i) \\
  \ldots\\
   \z_{N,i}^{\ast} \v_{N}(i) \\
\end{array}\right]
[ \v^{\ast}_{1}(i) \z_{1,i},\ldots, \v^{\ast}_{N}(i) \z_{N,i}]{\mathcal C}
\end{align}
The $(k,j)$-th block of the above matrix can be computed as:
\begin{equation}
 [{\mathcal G}]_{k,j}
=\left\{
\begin{array}{ll}
0,                                      & k \neq j \\
\sigma^2_{v,k}(R_{u,k}+\sigma^2_{n,k}I), & k=j
\end{array}
\right.
\end{equation}
and (\ref{eq.:def_G}) follows.
\section{computation of ${\Pi}$}
\label{apex.:mathcalN_calculation}
We rewrite ${\Pi}$ as:
\small
\begin{align}
{\Pi}=\E[\boldsymbol{\cal{P}}_i \Omega \boldsymbol{\cal{P}}^*_i]
\end{align}
\normalsize
where $\Omega=\omega^o \omega^{o*}$. The $(k,j)$-th block of ${\Pi}$ can be computed as:
\begin{align}
{\Pi}_{k,j}=&\E\sum_{\ell}\sum_m c_{\ell,k}c_{m,j}( \z_{\ell,i}^{\ast} \n_{\ell,i}-\sigma^2_{n,\ell}I)\nonumber \\
&\hspace{2cm}\times \Omega_{kj}( \n_{m,i}^{\ast}\z_{m,i}-\sigma^2_{n,m}I)
\end{align}
We use (\ref{eq.:network_linear_model_eq1}) to replace  $\z_{\ell,i}$ and $\z_{m,i}$:
\begin{align}
{\Pi}_{k,j}&=\E\sum_{\ell}\sum_m c_{\ell,k}c_{m,j}(\u_{\ell,i}^{\ast}\n_{\ell,i}+\n_{\ell,i}^{\ast}\n_{\ell,i}-\sigma^2_{n,\ell}I) \Omega_{kj}\nonumber \\
&\hspace{1cm}\times( \n_{m,i}^{\ast} \u_{m,i}+\n_{m,i}^{\ast} \n_{m,i}-\sigma^2_{n,m}I)
\end{align}
This leads to:
\begin{align}
{\Pi}_{k,j}=&\sum_{\ell}\sum_m c_{\ell,k}c_{m,j} \E[\u_{\ell,i}^{\ast}\n_{\ell,i} \Omega_{kj} \n_{m,i}^{\ast} \u_{m,i}]\\
&+\sum_{\ell}\sum_m c_{\ell,k}c_{m,j} \E[\n_{\ell,i}^{\ast}\n_{\ell,i} \Omega_{kj} \n^{\ast}_{m,i} \n_{m,i}^{\ast}]\nonumber \\
&-\sum_{\ell}\sum_m c_{\ell,k}c_{m,j} \E[\sigma^2_{n,\ell}  \Omega_{kj} \n^{\ast}_{m,i} \n_{m,i}]
\end{align}
If we assume that the regression $\{\u_{k,i}\}$ and noise $\{\n_{k,i}\}$ are zero mean circular Gaussian complex-valued vectors with uncorrelated entries, then:
\begin{equation}
\E[\u_{\ell,i}^{\ast}\n_{\ell,i} \Omega_{kj} \n_{m,i}^{\ast} \u_{m,i}]=
\begin{cases}
 0                     &  \ell\neq m \\
\sigma^2_{n,\ell} \Tr(\Omega_{kj}) R_{u,\ell} &  \ell= m
\end{cases}
\end{equation}
\begin{align}
&\E[\n_{\ell,i}^{\ast}\n_{\ell,i} \Omega_{kj} \n_{m,i}^{\ast} \n_{m,i}]= \nonumber \\
&
\begin{cases}
 \sigma^2_{n,\ell} \Omega_{kj}\sigma^2_{n,m}                                                     &  \ell \neq m \\
 \beta \sigma^2_{n,\ell} \Omega_{kj}\sigma^2_{n,m} +\sigma^2_{n,\ell} I\, \Tr(\Omega_{kj} \sigma^2_{n,\ell} I)    &  \ell= m
\end{cases}
\end{align}
and
\be
\E[\sigma^2_{n,\ell} I \Omega_{kj} \n^{\ast}_{m,i} \n_{m,i}]=\sigma^2_{n,\ell} \Omega_{kj}\sigma^2_{n,m}
\ee
We note that 
\be \Omega_{k,j}=w^o_k\, w^o_j{^*}
\ee 
where \mbox{$w^o_j=w^o_k,\; \forall \, k,j \in \{1,2,\ldots N\}$}. Therefore,
\begin{equation}
\Omega_{\ell k}=\Omega_{mn},\quad \forall \ell,k,m,n \in \{1,2,\cdots,N\}
\end{equation}
and $ \Tr(\Omega_{kj})=\|w^o\|^2$. As a result:
\begin{align}
{\Pi}_{k,j}=&\sum_{\ell} c_{\ell,k}c_{\ell,j}\Big\{\sigma^2_{n,\ell} \|w^o\|^2 \big(R_{u,\ell}+ \sigma^2_{n,\ell} I\big)\nonumber \\
&\hspace{3cm}+(\beta-1)\sigma^4_{n,\ell}w^o w^{o*}\Big\}
\end{align}

\end{appendices}
\section*{Acknowledgement}
The authors would like to thank Prof. Ali H. Sayed for his helpful suggestions
and advice during the initial phase of this work.

\typeout{}
\bibliographystyle{./IEEEtran}

\vspace{-1cm}
\begin{IEEEbiography}[{\includegraphics[width=1in,height=1.25in,clip,keepaspectratio]{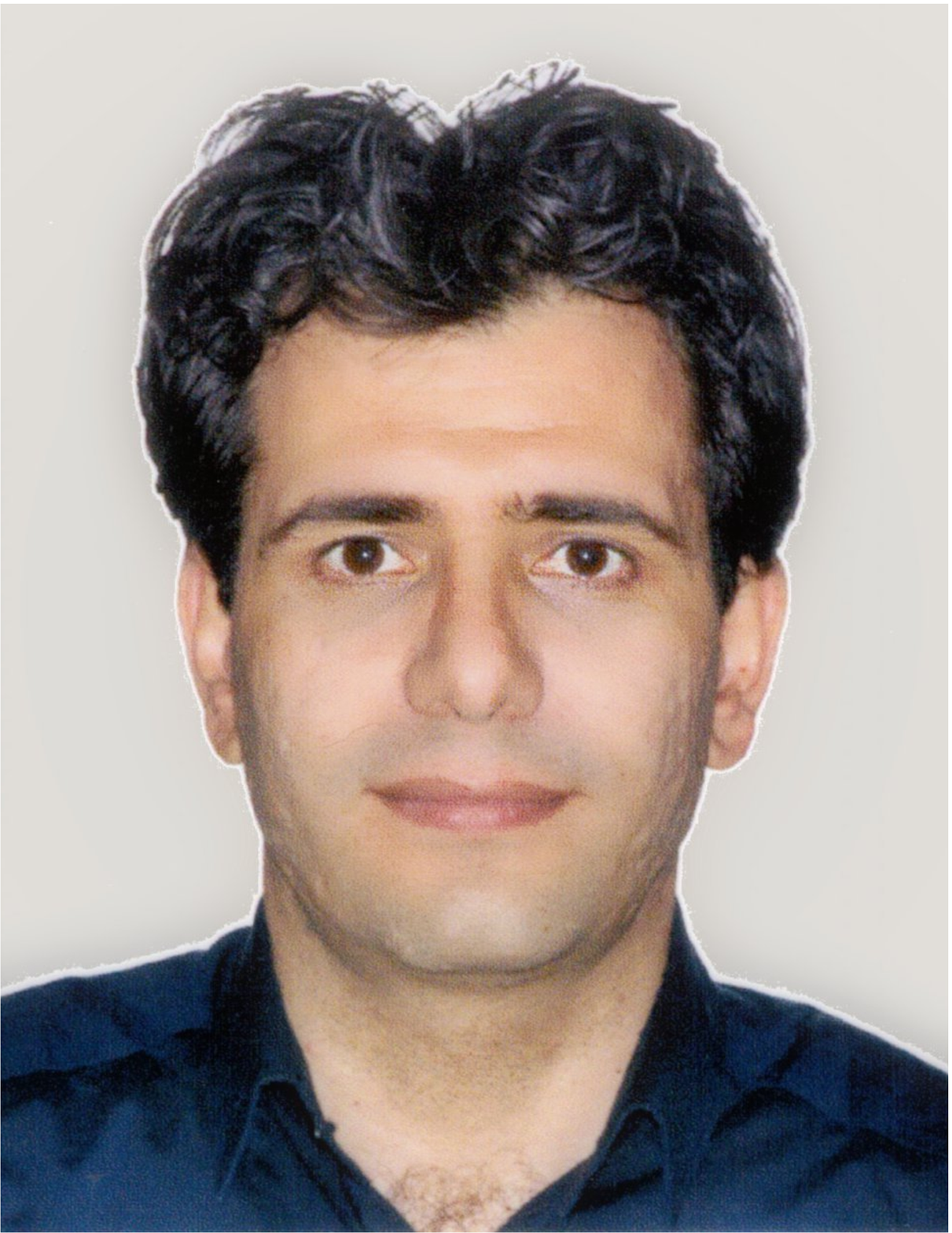}}]{Reza Abdolee} is currently a Ph.D. candidate at the Department of Electrical and Computer Engineering, McGill University, Montreal, Canada. In 2012, he was a research scholar at the Bell Labs, Alcatel-Lucent, Stuttgart, Germany. In 2011, he was a visiting Ph.D. student at the
Department of Electrical Engineering, University of California, Los Angeles (UCLA).  From 2006 to 2008, Mr. Abdolee worked as
a staff engineer at the Wireless Communication Center, University of Technology, Malaysia (UTM), where he implemented a switch-beam smart antenna system for wireless network applications. His research interests include communication theory, statistical signal processing, optimization, and hardware design and integration.  Mr. Abdolee was a recipient of several awards and scholarships, including, NSERC Postgraduate Scholarship, FQRNT Doctoral Research Scholarship, McGill Graduate Research Mobility Award, DAAD-RISE International Internship scholarship (Germany), FQRNT International Internship Scholarship, McGill Graduate Funding and Travel award, McGill International Doctoral Award, ReSMiQ International Doctoral Scholarship.
\end{IEEEbiography}
\vspace{-1cm}
\begin{IEEEbiography}[{\includegraphics[width=1in,height=1.25in,clip,keepaspectratio]{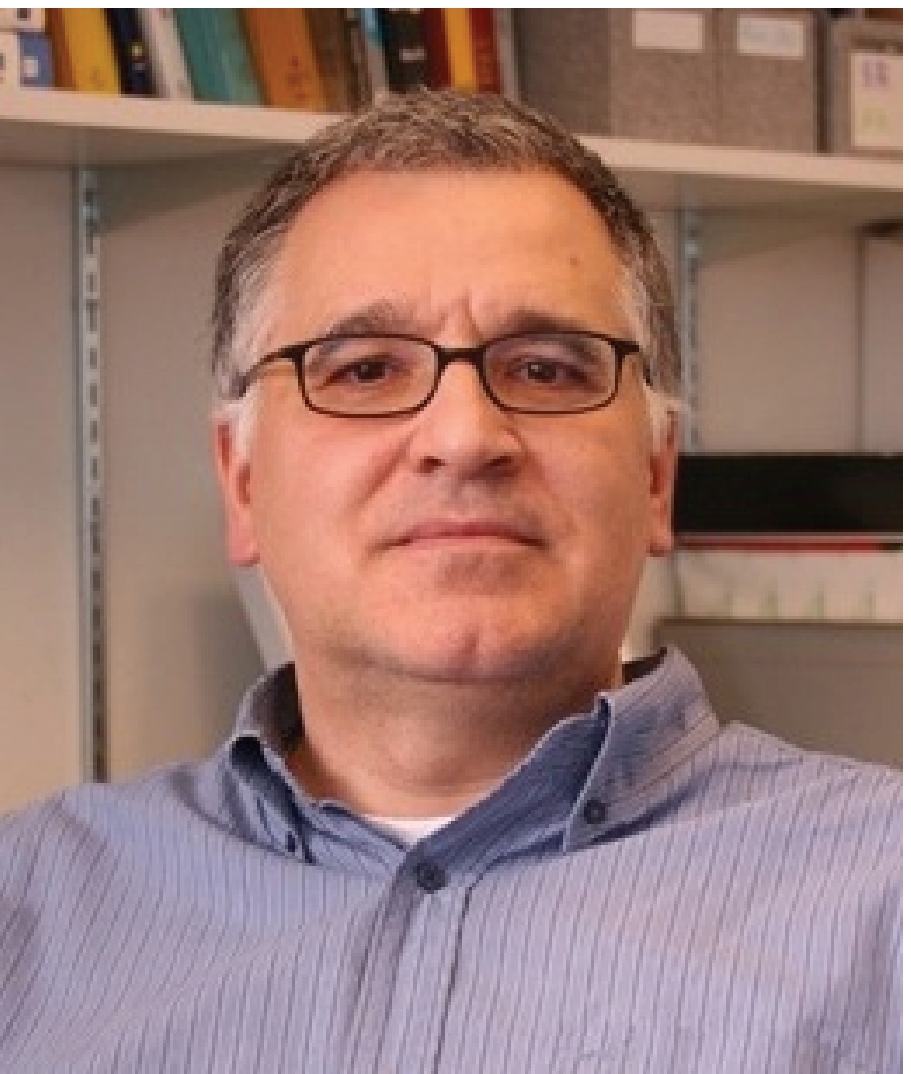}}]{Benoit Champagne} received the B.Ing. degree in Engineering Physics from the Ecole Polytechnique de Montréal in 1983, the M.Sc. degree in Physics from the Université de Montréal in 1985, and the Ph.D. degree in Electrical Engineering from the University of Toronto in 1990. From 1990 to 1999, he was an Assistant and then Associate Professor at INRS-Telecommunications, Université du Quebec, Montréal. In1999, he joined McGill University, Montreal, where he is now a Full Professor within the Department of Electrical and Computer Engineering. He also served as Associate Chairman of Graduate Studies in the Department from 2004 to 2007.

His research focuses on the development and performance analysis of advanced algorithms for the processing of information bearing signals by digital means. His interests span many areas of statistical signal processing, including detection and estimation, sensor array processing, adaptive filtering, and applications thereof to broadband communications and speech processing, where he has published extensively. His research has been funded by the Natural Sciences and Engineering Research Council (NSERC) of Canada, the Fonds de Recherche sur la Nature et les Technologies from the Government of Quebec, Prompt Quebec, as well as some major industrial sponsors, including Nortel Networks, Bell Canada, InterDigital and Microsemi.

\end{IEEEbiography}
%
\end{document}